\documentclass[oneside,a4paper,12pt,shownumbers,manualsort]{article}
\usepackage{latexsym}
\usepackage{euscript}
\usepackage{epsfig,amsmath,amssymb}

\renewcommand\({\left(}
\renewcommand\){\right)}
\renewcommand\[{\left[}
\renewcommand\]{\right]}

\newcommand{\dd}{{\rm d}}
\newcommand{\e}{{\rm e}}
\def\be{\begin{equation}}
\def\ee{\end{equation}}
\def\bea{\begin{eqnarray}}
\def\eea{\end{eqnarray}}

\newcommand\TeV{\,\mbox{TeV}}
\newcommand\GeV{\,\mbox{GeV}}

\newcommand\mpl{m_{\rm p}}
\newcommand\mcN{\mathcal N}

\newcommand\mcO{\mathcal O}
\long\def\symbolfootnote[#1]#2{\begingroup%
\def\thefootnote{\fnsymbol{footnote}}\footnote[#1]{#2}\endgroup}

 \large\normalsize

\begin{document}
%\draft
\begin{center}
{\Large   \bf                                                                   Confronting hybrid inflation in supergravity with CMB data
}

\vspace*{7mm}
{\ Rachel Jeannerot$^{a}$\symbolfootnote[1]{{E-mail:jeannerot@lorentz.leidenuniv.nl}} and Marieke Postma$^{b}$}\symbolfootnote[2]{E-mail:mpostma@nikhef.nl}
\vspace*{.25cm}

${}^{a)}${\it Instituut-Lorentz for Theoretical Physics,
Niels Bohrweg 2, 2333 CA Leiden, The Netherlands}\\
\vspace*{.1cm} 
${}^{b)}${\it NIKHEF, Kruislaan 409, 1098 SJ Amsterdam,
The Netherlands}

\begin{abstract}
  $F$-term GUT inflation coupled to $N=1$ Supergravity is confronted
  with CMB data. Corrections to the string mass-per-unit-length
  away from the Bogomolny limit are taken into account.  We find that
  a superpotential coupling $10^{-7}/\mcN \lesssim \kappa \lesssim
  10^{-2}/\mcN$, with $\mcN$ the dimension of the
  Higgs-representation, is still compatible with the data.  The
  parameter space is enlarged in warm inflation, as well as in the
  curvaton and inhomogeneous reheat scenario.  $F$-strings formed at
  the end of $P$-term inflation are also considered.  Because these
  strings satisfy the Bogomolny bound the bounds are stronger: the
  gauge coupling is constrained to the range $10^{-7} < g <10^{-4}$.
\end{abstract}
\end{center}

%\newpage

\section{Introduction}

The cosmic microwave background (CMB) power spectrum measured by WMAP
points to a predominantly adiabatic perturbation spectrum, as produced
in standard inflation \cite{WMAPinflation}. The existence of the
acoustic peaks excludes cosmic strings as the main source of
perturbations, although a 10\% contribution is not excluded
~\cite{Pogosian}. This has important implications for hybrid inflation
\cite{hybrid} because almost all particle physics models of hybrid
inflation, such as standard SUSY GUT $F$-term inflation, $D$-term
inflation and brane inflation, predict the formation of cosmic strings
at the end of inflation \cite{prd,jrs,polchinski}. The string
contribution to the CMB anisotropies in $D$-term models is far too
high \cite{prd} unless the gauge coupling constant is unnaturally
small \cite{Pterm}. In this paper we focus on $F$-term inflation
which can naturally arise from SUSY GUTs and as a low energy effective
description of a certain class of interacting $D$-brane models.

The minimal field content of $F$-term inflation is a gauge singlet
superfield and two Higgs superfields which transform in complex
conjugate representations of some gauge group $G$. Inflation takes
place as the singlet field slowly rolls down a valley of local minima
where the vacuum expectation value (VEV) of the Higgs fields vanish
and $G$ is unbroken. When the scalar singlet falls below a certain
critical value, the Higgs mass becomes tachyonic and inflation ends
very rapidly, via a phase transition during which the Higgs fields
acquire a non-vanishing VEV and $G$ spontaneously breaks down to a
subgroup $H$.  Of all topological defects that can form during such a
phase transition, only cosmic strings are cosmologically viable.

%Neither monopoles nor domain walls should form during
%this phase transition. In the context of GUTs where monopoles ought to
%form, they must be formed during a previous phase transition. 
%It is therefore not
%the breaking the GUT gauge group $G$ which takes place at the end of
%inflation but the breaking of an intermediate symmetry group. 

In the context of GUTs, where monopoles ought to form, this implies
that monopoles should be formed in a phase transition prior to
inflation.  Hence, the group $G$ which gets broken at the end of
inflation is not the GUT group itself, but some intermediate symmetry
group. There will be more phase transitions after inflation if $H$ is
larger than the standard model gauge group, and these should not lead
to the formation of unwanted defects. These arguments together with
the observation that the rank of the symmetry group $G$ which is
broken at the end of inflation is reduced by one unit (${\rm rank }
(G) = {\rm rank}(H) +1$) and simple homotopy arguments, lead to the
conjecture that cosmic strings always form at the end of GUT hybrid
inflation \cite{prd}. This was proven for an SO(10) GUT \cite{prd1},
and recently for all GUT groups with rank less than 8 \cite{jrs}. The
strings which form at the end of inflation are topological and remain
stable down to low energies if $R$-parity remains unbroken.

The constraints from the CMB data can be avoided if the cosmic strings
are unstable, or if no strings are formed at all. We note that this
only happens in rather specific models. The embedded strings arising
in GUT models in which $R$-parity is broken can be unstable, depending
on the particular model.  Strings can be made semi-local, and thus
unstable, by adding extra charged chiral multiplets \cite{ana2}, or by
assuming that the Higgs fields also transform non-trivially under some
non-Abelian group \cite{kallosh}. If the gauge symmetry is already
broken during inflation, there are no strings at all. This can be done
by either adding a non-renormalisable term in the superpotential
\cite{shifted}, extra GUT Higgs superfields \cite{shifted2}, or a
discrete symmetry \cite{smooth}.

In this paper we determine the parameter range for which standard
$F$-term inflation is compatible with CMB data.  The strings that form
at the end of GUT $F$-inflation do not satisfy the Bogomolny bound.
Therefore the string tension is a function of the Higgs and gauge
field masses. This is in contrast with the $F$-term inflation models
that emerge as an effective description of brane inflation
\cite{Pterm}, as the strings formed in these models do satisfy the
Bogomolny bound.  Note that SUSY is broken in the core of the strings
and hence there are fermionic zero modes solutions bounded to the
strings in the global SUSY case \cite{acd}. These zero modes disappear
when gravity is included \cite{sugrastr}. $F$-strings do not carry any
current and hence the tension and energy-per-unit-length of the strings
are equal \cite{sugrastr,maj}. Our work improves in two ways on the
existing literature~\cite{covi,shafi1,Rocher,mar}.  First of all,
for GUT strings we take the corrections to the string tension away
from the Bogomolny limit into account.  This enlarges the parameter
space.  Secondly, we include in a consistent manner all Supergravity
corrections to the potential.  To do so we assume general hidden
sector supersymmetry breaking.  Possible dissipative corrections are
also discussed.

The layout of this paper is as follows.  In the next section we
introduce the potential of standard $F$-term hybrid inflation, and
include all SUSY breaking and SUGRA corrections.  In section
\ref{s:CMB} we address the bounds on the symmetry breaking scale
implied by the data.  Apart from the ``10\%-bound'' mentioned at the
very beginning of this introduction, we also give the Kaiser-Stebbins
bound, and the bound coming from pulsar observations. The density
perturbations produced by hybrid inflation are discussed in section
\ref{s:perturbations}.  Setting them equal to the observed spectrum
gives the symmetry breaking scale $M$ as function of the
Higgs-inflaton coupling $\kappa$.  This allows to translate the
various bounds on $M$ in bounds on $\kappa$.  We derive an analytic
expression for $M(\kappa)$ in the limit where one term dominates the
potential.  However, our approximation breaks down in the large
coupling limit and numerical calculations are needed.  Our numerical
results are presented in section \ref{s:numerical}.  We discuss both
the constraints on GUT and on brane $F$-term inflation.  Finally, in
section \ref{s:warm} we discuss warm inflation, occurring when the the
inflaton or Higgs fields can decay during inflation and the
corresponding dissipative terms are important.  Dissipation can
ameliorate the CMB bounds.

\section{Hybrid inflation --- The potential}
\label{s:potential}

The superpotential for standard hybrid inflation is given by
\cite{Cop,Dvasha}
\be 
W_{\rm inf} = \kappa S( \phi_+ \phi_{-} - M^2),
\label{W}
\ee
with $S$ a gauge singlet superfield, and $\phi_+$, $\phi_-$ Higgs
superfields in complex conjugate representations of a gauge group G.
In this paper, we use the same notation for the superfield and their
scalar components. The coupling constant $\kappa$ and the symmetry
breaking scale $M$ can be taken real and positive without loss of
generality. The supersymmetric part of the scalar potential is given
by
\bea V_{\rm SUSY} &=& \sum_b \left| \frac{\partial W}{\partial \phi_b}
\right|^2 + \frac{g^2}{2} \sum_{a} \(\sum_{b} \phi_{b,i}^\dagger
t^{a,i}_j \phi_b^j\)^2
\nonumber \\
&=& \kappa^2 |\phi_+ \phi_- - M^2|^2 + \kappa^2
|S|^2(|\phi_+|^2+|\phi_-|^2) + V_D
\label{Vsusy}
\eea
where the sum $b$ is over all fields. $t^a_i$ are the generators of
$G$, $a=1...n$ where $n$ is the dimension of $G$, and $i,j = 1...\mcN$
where $\mcN$ is the dimension of the representation of the field
$\phi_b$.  In Eq.(\ref{Vsusy}) $\phi_-$ and $\phi_+$ refer to the
scalar components of the corresponding superfields which acquire a
VEV after inflation.  Vanishing of the $D$-terms enforces $|\phi_-| =
|\phi_+|$.  Assuming chaotic initial conditions the fields get trapped
in the inflationary valley of local minima at $|S| > S_c = M$ and
$\phi_- = \phi_+ = 0$, where $G$ is unbroken. The potential is
dominated by a constant term
\be
V_0 = \kappa^2 M^4
\label{V_0}
\ee
which drives inflation.  Inflation ends when the inflaton drops below
its critical value $S_c$ (or when the second slow-roll parameter
$\eta$ equals unity, whatever happens first) and the fields roll
toward the global SUSY minima of the potential $|\phi_+| = |\phi_-| =
M$ and $S=0$.  During this phase transition the gauge group $G$ is
spontaneously broken down to a subgroup $H$.  Cosmic strings form via
the Kibble mechanism if the vacuum manifold $G/H$ is simply connected
\cite{Kibble}.  If $G$ is embedded in a GUT theory, or $G = U(1)$ as
is the case in effective $D$-brane models, cosmic strings
form~\cite{prd,jrs}.

In the standard scenario the flatness of the tree level potential is
lifted by loop corrections \cite{Dvasha}.  These do not vanish during
inflation because $F_S \neq 0$ and SUSY is broken. The two scalar mass
eigenstates $\chi_\pm = 1/\sqrt{2}(\phi_+ \pm \phi_-)$ have masses
$m_\pm^2=\kappa^2(S^2 \pm M^2)$, while their fermionic superpartners
both have mass $\tilde{m}_\pm^2 = \kappa^2 S^2$. If the Higgs
representation is $\mcN$-dimensional, there are $\mcN$ such
mass-splitted double-pairs. The one loop correction to the potential
can be calculated using the Coleman-Weinberg formula \cite{CW} $V_{\rm
loop} = \frac{1}{64 \pi^2} \sum_i(-)^{F_i}\ M_i^4 \ln
\frac{M_i^2}{\Lambda^2}$, which for the superpotential in
Eq.~(\ref{W}) gives\footnote{When $|S|$ is very close to $\mpl$ there
are SUGRA corrections to the masses of the scalars and fermions which
enter the loop correction. However, as we shall see further, the loop
corrections dominate when $|S|$ is very close to $S_c \ll \mpl$, and
these corrections do no play any r\^ole.}  \cite{Dvasha}\\
\be
V_{\rm loop} 
= \frac{\kappa^4 M^4 \mcN}{32 \pi^2} \[
2 \ln\(\frac{M^2 \kappa^2 z }{\Lambda^2}\) + (z+1)^2 \ln(1+z^{-1}) +
(z-1)^2 \ln (1-z^{-1}) \]
\label{V_loop}
\ee
with
\be
z = x^2 = \frac{|S|^2}{M^2}.
\ee

\subsection{SUGRA corrections}

In addition to $V_{\rm loop}$ there are SUGRA corrections to the
potential, i.e., corrections that vanish in the limit that the Planck
mass is taken to infinity and gravity decouples.  In any model that
aspires to describe the real world, there are two sources of SUSY
breaking: SUSY breaking by the finite energy density during inflation
and SUSY breaking in the true vacuum; the later is responsible for the
soft terms today. Both sources of breaking contribute to the SUGRA
corrections.  These corrections therefore depend on the particular
scenario for SUSY breaking at low energy and in particular on the
form of the hidden sector superpotential.

The superpotential gets a contribution from both the inflaton and
hidden sector potential $W_{\rm tot} = W_{\rm inf}(S,\phi_+,\phi_-) +
W_{\rm hid}(z)$.  In gauge mediated SUSY breaking models there is also
a contribution from the messenger sector and in general GUT models
from other GUT superfields. We assume that they do not couple to the
inflaton sector except gravitationnaly. The hidden sector expectation
values at the minimum of $V$ may generically be written as
\be
\langle z \rangle = a \mpl,
\qquad
\langle W_{\rm hid} \rangle = \mu \mpl^2,
\qquad
\langle \frac{\partial W_{\rm hid}}{\partial z} \rangle = c \mu \mpl,
\ee
with $\mpl = (8\pi G)^{-1/2} = 2.4 \times 10^{18} \GeV$ the reduced
Planck mass, $a,c$ dimensionless numbers, and $\mu$ a mass parameter
characterizing the VEV of the hidden-sector superpotential.  Setting
the cosmological constant to zero by hand ($\langle V \rangle = 0$
after inflation) requires tuning
\be
|c + a^*|^2 = 3.
\label{ac}
\ee
The scalar potential is
\be
V = \e^{K/\mpl^2} \left[ \sum_\alpha
\Big| \frac{\partial W}{\partial \phi_\alpha} 
+ \frac{\phi_\alpha^* W}{\mpl^2} \Big|^2
- 3 \frac{|W|^2}{\mpl^2}
\right]
\label{Vsugra}
\ee
where the sum is over all fields.  We take minimal kinetic terms,
corresponding to a K\"ahler potential $K = \sum_\alpha
|\phi_\alpha|^2$.  The true vacuum gravitino mass is then given by
\be
m_{3/2} = \e^{|a|^2/2} \mu.
\ee
In gravity mediated SUSY breaking schemes $m_{3/2} \sim \TeV$ is of
the order of the soft mass terms, whereas it can be smaller in
gauge mediated schemes.

During inflation when $|S|>S_c$ and the SUGRA $F$-term $F_\alpha =
\partial W_{\rm tot}/\partial \phi_\alpha + \frac{\phi_\alpha^* W_{\rm
    tot}}{\mpl^2} \neq 0$ for $\phi_\alpha = z,S$. This is the SUGRA
generalization of $F$-term SUSY breaking.   Following
Ref.~\cite{martin}, we rescale the visible sector superpotential
$W_{\rm inf} \to \e^{-|a|^2/2} W_{\rm inf}$ in order to recover from
Eq.~({\ref{Vsugra}) the properly normalized tree level potential in
the global limit given by Eq.~(\ref{Vsusy}), i.e, in the limit when $\mpl \to
  \infty$ and gravity decouples.

Expanding the exponential term in Eq.~(\ref{Vsugra}) in powers of
$|S|/ \mpl$, we find the SUGRA corrections to $V_{\rm SUSY}$
\footnote{We neglect higher order corrections to the K\"ahler and
super potential; we expect these terms to change the coefficients in
front of the various terms but not their qualitative structure. They
could also destabilise the VEV of the Higgs fields and we assume here
that they do not.}
\begin{equation}
V_{\rm SUGRA} = V_{\rm A} + V_{\rm m} + V_{\rm NR}.
\end{equation}
The A-terms are of the form
\be 
V_A = 2 \kappa M^2 m_{3/2} |S| \cos (\arg \mu - \arg S) \(2 +
\frac{|S|^2}{\mpl^2} + ...\) \ee
with the ellipsis denoting higher powers of $|S|/\mpl$.  The
linear term is dominant. It is proportional to both the high and low
energy SUSY breaking scale.  This term was discussed in \cite{covi} in
the context of an explicit O'Raifeartaigh model, and more recently 
in \cite{shafi1}.

The mass term is of the form
\be
V_m = \( 3 H^2 (|a|^2 + ...)  - 2 m_{3/2}^2 \)|S|^2
\label{Vm}
\ee
where we have expanded the $\e^{|a|^2}$ in powers of $|a|$, which is
only valid for $|a| \ll 1$; the ellipsis denote higher order terms in
$|a|$. We will refer to the first term as the Hubble induced mass. The
second term, the vacuum soft mass term, is negligible small compared to
the other SUGRA contributions.  The Hubble parameter during inflation
is given by the Friedman equation: $H^2 =V/(3\mpl^2) \approx V_0
/(3\mpl^2)$.  There is no soft mass in the absence of low energy SUSY
breaking ($a \to 0, m_{3/2} \to 0$).  This is a consequence of taking
minimal K\"ahler, the zeroth order term $V_{m} \sim H^2 |S|^2$
cancels; this can be considered as fine tuning.

A scale invariant perturbation spectrum in agreement with observations
is obtained for sufficiently small slow roll parameter $|\eta| \sim
|m_S^2|/H^2 \lesssim 10^{-2}$ \cite{WMAP}, see Eq.~({\ref{slow_roll})
below.  For generic K\"ahler potential $\eta \sim 1$, which is the
infamous $\eta$-problem.  Scale invariance thus requires minimal K\"ahler
potential (or close to it) to cancel to the zeroth order Hubble
induced mass term proportional to $|a|^0$, and $|a| \lesssim 0.1$ to
cancel the higher order terms proportional to $|a|^{2n}$ as given in
Eq.~(\ref{Vm}). This last requirement excludes the simplest Polonyi
model $W_{\rm hid} = M_s^2(\beta +z)$ with $z = \mcO
(1)\mpl$.~\footnote{The Hubble induced mass term in Eq.~(\ref{Vm})
comes from an expansion of $\e^{K/\mpl^2}$ in Eq.~(\ref{Vsugra}) to
second order in the fields; this term is missed in \cite{covi}.}

The non-renormalisable SUGRA terms are of the form
\be V_{NR} = {1\over 2} ({3}H^2 - m_{3/2}^2) |S|^2 \(
\frac{|S|^2}{\mpl^2} + ...\) \ee
with the ellipsis denoting higher order terms in $|S|^2/\mpl^2$.  The
Hubble induced term dominates. They have been discussed before
\cite{sugra}.

How generic are the above SUGRA corrections?  We assume minimal
K\"ahler potential and general hidden sector SUSY breaking.  A
non-minimal K\"ahler potential would generically be catastrophic, as
it means a large Hubble induced mass impeding slow roll inflation.
The hidden sector is characterized by the scale $m_{3/2}$ and the
dimensionless constant $|a|$. The A-terms are small for a small
gravitino mass, as can occur in gauge mediated SUSY breaking where
$m_{3/2} \ll \TeV$ is possible.  The Hubble induced mass term does not
depend on the gravitino mass; its effect can only be decreased by
taking $a \ll 1$.  Note that this entails a large hierarchy between
the dimensionless hidden sector parameters $a$ and $c$, see
Eq.~(\ref{ac}), implying some sort of fine tuning.  The
non-renormalisable term is independent of the SUSY breaking sector,
and is generic. If SUSY breaking in the true vacuum occurs after
inflation, i.e., $z$ only acquires its VEV after inflation, then
$\mu,\;a \to 0$: the mass and $A$-terms are absent, but the NR terms
are still there.\\

\noindent
The scalar potential including all corrections then is 
\be
V = V_{\rm SUSY} + V_{\rm loop} + V_{\rm SUGRA}.
\ee
During inflation $V_{\rm SUSY} \approx V_0 =\kappa^2 M^4$, and the
potential reads
\begin{eqnarray}
V &=& 
\kappa^2 M^4 \bigg[ 1 + \frac{\kappa^2  \mcN}{32 \pi^2} \Big[
2 \ln\(\frac{\kappa^2 |S^2| }{\Lambda^2}\) + (z+1)^2 \ln(1+z^{-1}) 
\nonumber\\ 
&+&
(z-1)^2 \ln (1-z^{-1}) \Big] +  {1\over 2}  {|S|^2\over m_p^2} \Big(
\frac{|S|^2}{\mpl^2} + ...\Big) 
+  \Big(|a|^2 + ...\Big) {|S|^2\over m_p^2} \bigg] 
\nonumber \\
&+& 2 
\kappa M^2 m_{3/2} |S| \cos (\arg \mu - \arg S) \Big(2 +
\frac{|S|^2}{\mpl^2} + ...\Big),
\end{eqnarray}
where we have omitted in $V_{\rm NR}$ and $V_{\rm m}$ the gravitino
mass dependent terms which are negligible during inflation in the
parameter range of interest.  Keeping only the dominant terms and
expressing everything in terms of the real inflaton field $\sigma =
\sqrt{2}|S|$ gives
\begin{eqnarray}
V 
&=& 
\kappa^2 M^4 \bigg[ 1 + \frac{\kappa^2  \mcN}{32 \pi^2} \Big[
2 \ln\(\frac{2 \kappa^2 \sigma^2}{\Lambda^2}\) + (z+1)^2 \ln(1+z^{-1}) 
\nonumber \\ 
&+&
(z-1)^2 \ln (1-z^{-1}) \Big] + {\sigma^4\over 8 m_p^4}  
+  {|a|^2 \sigma^2\over 2 m_p^2} \bigg] 
\nonumber\\
&+& 
\kappa  A m_{3/2} M^2  \sigma 
\end{eqnarray}
where $A= 2 \sqrt{2} \cos(\arg \mu - \arg S)$. Here we have assumed
that $\arg S$ is constant during inflation. Further, $z= x^2 =
{|S|^2}/M^2 = \sigma^2 /(2M^2)$ so that $z=x=1$ when $\sigma =
\sigma_c$.

%$V \approx V_0 = \kappa^2 M^4$

%$V_{\rm loop} = \frac{\kappa^4 M^4 \mcN}{32 \pi^2} \[
%2 \ln\(\frac{M^2 \kappa^2 z }{\Lambda^2}\) + (z+1)^2 \ln(1+z^{-1}) +
%(z-1)^2 \ln (1-z^{-1}) \]$

%$V_A = 2 \kappa M^2 m_{3/2} |S| \cos (\arg \mu - \arg S) \(2 +
%\frac{|S|^2}{\mpl^2} + ...\)$

%$V_m = \( 3 H^2 (|a|^2 + ...)  - 2 m_{3/2}^2 \)|S|^2$

%$ V_{NR} = {1\over 2} ({3}H^2 - m_{3/2}^2) |S|^2 \(
%\frac{|S|^2}{\mpl^2} + ...\)$

\section{CMB constraints}
\label{s:CMB}

\subsection{String tension}

SUSY is broken in the core of the strings which form at the end of
SUSY inflation and hence there are fermionic zero modes solutions
bounded to the strings in the global SUSY case \cite{acd}.  These zero
modes disappear when gravity is included \cite{sugrastr}.  $F$-strings
do not carry any current and hence the tension and energy-per-unit
length of the strings are equal \cite{sugrastr,maj}.  

Cosmic strings satisfying the Bogomolny bound have a tension $\mu = 2
\pi M^2$, with $M$ the VEV of the string Higgs fields far away from
the string.  The strings forming at the end of $F$-term inflation do
not satisfy this bound, and there are corrections to the simple
formula above, which depend on the ratio of the common Higgs mass
$m_\phi$ to the string's gauge boson mass $m_A$ \cite{Hill,ShelVil}
\be 
\mu = 2 \pi M^2 \epsilon(\beta) 
\label{mu}
\ee
where $\beta = (m_\phi / m_A)^2$. $m_\phi = \kappa M$ with $\kappa$
the superpotential coupling constant and $m_A$ is given in terms of
gauge coupling constant $g$; the exact relation depends on the
dimension and on the transformation properties of the representations
of the Higgs fields $\phi_+$ and $\phi_-$. $m_A = \sqrt{2} g M$ for
$\mcN =1$. In GUT models $g^2  \simeq 4\pi/25$ and $m_A \simeq
M$. In this paper we shall use $m_A \simeq 1$ unless stated otherwise.
In the Bogomolny limit $\epsilon(1) = 1$.  From Ref.~\cite{Hill}
\be
\epsilon(\beta) \approx \left \{
\begin{array}{lll}
1.04  \beta^{0.195}, & \qquad \beta > 10^{-2}, \\
{\displaystyle \frac{2.4}{ \log(2/\beta)}}, & \qquad \beta < 10^{-2}.
\end{array}
\right.
\label{epsilon}
\ee
For $m_A \simeq M$,  $\beta$ varies from $1$ to $10^{-12}$ as
$\kappa$ goes from $1$ to $10^{-6}$, and thus $\mu$ changes by a
factor $\sim 20$.  In Fig.~\ref{F:eps} we plotted $\epsilon(\beta)$ as
function of $\kappa$ for $m_A = M$.  For $\kappa \sim 10^{-2}$ the
bound on $\mu$, discussed below, is weaker by a factor $4$ if
$\epsilon(\beta)$ is taken into account.

\begin{figure}
\begin{center}
\leavevmode\epsfysize=8cm \epsfbox{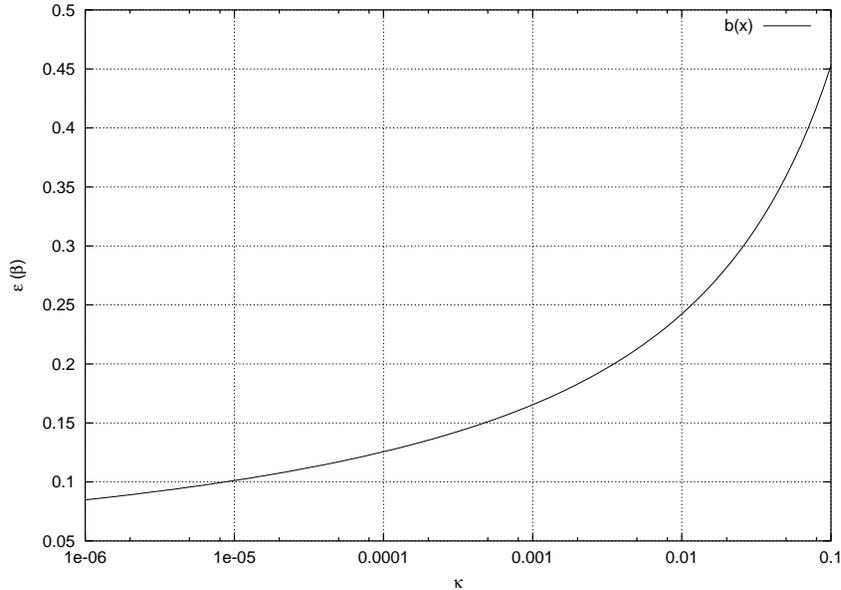}
\caption{ $\epsilon(\beta)$ vs. $\kappa$ for $m_A \simeq M$.}
\label{F:eps}
\end{center}
\end{figure}

\subsection{Bounds from CMB}

Cosmological perturbations from inflation and strings in hybrid models
are proportional to the same scale $M$. They are uncorrelated and thus
the multipole moments of the CMB power spectrum just add up
\cite{prd}. The proportionality coefficients depend on the GUT
parameters and on the normalised contribution of each component
\cite{prd}. Cosmic strings do not predict the acoustic peaks which
have been measured in CMB experiments and hence their contribution
should be rather small, less than 10\% at the $3\sigma$ level
\cite{Pogosian}. The contribution from cosmic strings depends on the
mass-per-unit-length $\mu$ and on the density properties of the string
network at last scattering. No full field theoretic simulations exist
and hence no robust prediction of the full power spectrum can be made
right now.  The model parameters can be constrained, but only up to
the uncertainties of the simulations.

The cosmic string contribution to the quadrupole is given by
\be
\( \frac{\delta T}{T} \)_{\rm cs} = y G \mu;
\label{y}
\ee
the parameter $y$ depends on the simulation. Recent work predicts
$y=8.9$ \cite{Landriau}.  The error margin they quote gives a range $y
= 6.7 - 11.6$.  Older simulations give $y = 6$~\cite{Allen}, and
semi-analytic approximations give $y = 3 -6$~\cite{approx}.

%Normalising the string contribution to CMB data gives an upper bound
%on $G\mu$. 

The quadrupole measured by COBE (which coincides with WMAP
data) is \cite{WMAP,COBE}
\be \(\frac{\delta T}{T}\)_{\rm COBE} = 6.6 \times 10^{-6}.
\label{cobe}
\ee
%h
The cosmic string contribution to the quadrupole is
given by
\be 
B = \Big| \frac{(\delta T/T)_{\rm cs}} 
{({\delta T}/{T})_{\rm COBE}} \Big|^2.  
\label{B}
\ee
The analysis of Ref.~\cite{Pogosian} gives the bound $B < 0.1$, i.e.,
a string contribution less than 10\%.  Using Eqs.({\ref{y}) and
(\ref{mu}) this implies 
\be
G \mu < 6.9 \times 10^{-7} \( \frac{3}{y} \) 
\quad \Rightarrow \quad
M_{str} < 4.1 \times 10^{15} 
\sqrt{\frac{(3/y)}{\epsilon(\beta)}}.
\label{Pogosian}
\ee
In our numerical simulations we will use the conservative value $y=3$.
The bound on $M$ is a factor $\sqrt{3}$ stronger for $y =9$, the value
suggested by the most recent simulations.

If there are extra dimensions the string reconnection probability $p$
(the probability that two strings reconnect when they pass through
each other), which is one for four dimensional gauge theory solitons,
can be less than unity~\cite{polchinski,dvali}. The result is that more
energy is stored in the string network at a given time, and thus the
constraint on the string scale in Eq.~(\ref{Pogosian}) becomes tighter
by a factor $\sqrt{p}$ \cite{Pogosian}.

Strings can also influence the CMB pattern after the time of last
scattering, through the Kaiser-Stebbins effect~\cite{stebbins}.  If a
moving string is traversing the photons coming towards us this will
give rise to a step-like discontinuity on small angular scales, the
step size being proportional to the mass-per-unit-length of the
string. The absence of such a discontinuity gives the bound \cite{Smoot} 
\be G \mu < 3.3 \times 10^{-7} \quad \Rightarrow \quad M < \frac{2.8
  \times 10^{15}}{ \sqrt{\epsilon(\beta)}}
\label{smoot}
\ee
The Kaiser-Stebbins(KS)-bound is stronger than the 10\%-bound for $y <
6.3$.  The KS-bound depends less on the evolution of the
string network, i.e., on the results of numerical simulations; in this
sense it is a much stronger bound.

Finally, the string scale can be bounded by constraining the
stochastic gravitational wave background produced by the string
network~\cite{ShelVil}.  Such a background would distort the regularity
of pulsar timing. No such distortion has been observed~\cite{Lommen},
which translates to
\be
G \mu < 1 \times10^{-7}
\quad \Rightarrow \quad
M <  \frac{1.5 \times 10^{15}}{\sqrt{\epsilon(\beta)}}. \label{pulsar}
\ee
Although the pulsar-bound is the most stringent of the three, it is
also the one with the largest uncertainties.  Due to the huge range of
scales involved, one of the least certain aspects of cosmic string
dynamics is the long term evolution of small scale structure.  And it
is this small scale structure that governs the gravitational
radiation.  

In our simulations, we shall use the three different bounds given in
Eqs.({\ref{Pogosian}), (\ref{smoot}) and (\ref{pulsar}).  In our
analytic formulas we will use $M < M_{\rm CMB} \approx 3 \times
10^{15} \GeV$, which corresponds to the KS-bound.

Before closing this section, we would like to mention that a lensing
event consistent with a cosmic string has been reported recently
\cite{sazhin}.  Further observation is needed to determine whether the
lensing is indeed string induced. Such a string would have
\be
G \mu \simeq 4 \times10^{-7}
\quad \Rightarrow \quad
M \simeq  \frac{3 \times 10^{15}}{\sqrt{\epsilon(\beta)}}.
\ee
to produce the observed image separation. If it is indeed a cosmic
string, we are on the verge of seeing it in the CMB data.

\section{Density perturbations}
\label{s:perturbations}

We recall here the important formulae for the density perturbations
\cite{LythLid}. The number of e-foldings before the end of inflation
is
\be N_Q = \int_{\sigma_{\rm end}}^{\sigma_Q} \frac{1}{\mpl^2}
\frac{V}{V'} \dd \sigma
\label{N_Q}
\ee
where the prime denotes derivative with respect to the normalised real
scalar field $\sigma \equiv \sqrt{2}|S|$, and the subscript $Q$
denotes the time observable scales leave the horizon, which happens
$N_Q \approx 60$ $e$-folds before the end of inflation.  $\sigma_{\rm
end}$ is the inflaton VEV when inflation ends, which is either the
critical point where the Higgs mass becomes tachyonic $\sigma_{\rm
end} = \sigma_c = \sqrt{2}M$, or the value for which one of the slow
roll parameters
\be
\epsilon = \frac 12 \mpl^2 \(\frac{V'}{V}\)^2,
\qquad \qquad
\eta = \mpl^2 \(\frac{V''}{V}\),
\label{slow_roll}
\ee
exceeds unity.  For hybrid inflation $\epsilon \ll \eta$ and can be
neglected.  The second slow roll parameter $\eta$ blows up in the
limit $x \to 1$ (as a consequence of the field $\chi_- = (\phi_+ -
\phi_-)/\sqrt{2}$ becoming massless), and thus determines the end of
inflation.  As we will see, for small enough coupling $\sigma_{\rm
end} \approx \sigma_c$. The inflaton contribution to the CMB
quadrupole anisotropy is
\be
\( \frac{\delta T }{T} \)_{\rm inf}
%= \frac{1}{12 \sqrt{10} \pi \mpl^2} \sqrt{\frac{V}{\epsilon}}
= \frac{1}{12\sqrt{5} \pi \mpl^3} \frac{V^{3/2}}{V'},
\label{dTphi}
\ee
evaluated at $\sigma = \sigma_Q$.  The tensor contribution $( {\delta
T}/{T} )_{\rm tens} \approx 0.03 H_*/\mpl $ is small, and can be
neglected.  The total anisotropy is
\be
\( \frac{\delta T }{T} \) =
\sqrt{
\( \frac{\delta T }{T} \)_{\rm inf}^2
+ \( \frac{\delta T }{T} \)_{\rm cs}^2+ 
 \( \frac{\delta T }{T} \)_{\rm s}
} 
\label{dTtot}
\ee
Here we have included a possible contribution from a scalar field
different from the inflaton, denoted by $(\delta T/T)_{\rm s}$.  This
term can be important if alternative mechanisms for density
perturbation are at work, as is the case in the
curvaton~\cite{curvaton} and inhomogeneous reheat scenario~\cite{irs}.
In these scenarios the fluctuations of the curvaton field respectively
the field modulating the inflaton decay rate gives the dominant
contribution to the density perturbations.  We define
\be
\delta_C = 
\frac{ (\delta T /T)_{\rm inf} + (\delta T /T)_{\rm cs}}
{(\delta T /T)_{\rm COBE}}.
\label{deltaC}
\ee
If the inflaton sector including cosmic strings is the only source of
perturbations $\delta_C =1$, whereas $\delta_C \ll 1$ in the curvaton
and inhomogeneous reheat scenario.  Note that the curvaton scenario in
its simplest form can only work for $H \sim \sqrt{V_0 / \mpl^2}
\gtrsim 10^7 \GeV$~\cite{lyth,postma}.

Finally, the spectral index is given by
\be
n_s - 1 = - 6 \epsilon + 2 \eta
\label{spectral}
\ee
WMAP bounds $n_s = 0.99 \pm 0.04$~\cite{WMAP}.

The potential energy $V_0$ dominates and drives inflation.  The
amplitude of the density perturbations is set by $V'$.  The spectral
index is determined by the second derivative $V''$.  The strategy to
compute the perturbation spectrum is the following. First determine
the inflaton VEV when inflation ends from the condition
$|\eta(\sigma_{\rm end})| =1$.  Then use Eq.~(\ref{N_Q}) to find the
inflaton VEV when observable scales leave the horizon $\sigma_Q$.
Finally, compute the quadrupole using Eqs.~(\ref{dTphi},\ref{dTtot})
evaluated at $\sigma = \sigma_Q$.  Setting the quadrupole equal to the
observed COBE value gives the symmetry breaking scale $M$ as a
function of the superpotential coupling $\kappa$.

In various regions of the parameter space, various corrections
dominate.  An analytic approximation to the perturbation spectrum is
then possible in the small coupling limit (where $x_{end} \approx x_Q
\approx 1$), as we will discuss now.

\subsection{The loop regime}

Assume first that the loop potential $V_{\rm loop}$ dominates.  The
derivative of the potential, which is needed for both the
determination of $N_Q$ and $\delta T/T$, is \cite{Dvasha}
\be
V'_{\rm loop} = \frac{\kappa^4 M^3 \mcN}{8 \sqrt{2} \pi^2} x f(x^2),
\ee
with
\be
f(z) = (z+1) \ln (1+z^{-1}) + (z-1) \ln (1-z^{-1}).
\ee
Recall that in hybrid inflation the first slow roll parameter is
negligible with respect to the second $\epsilon \ll \eta$. To
determine the end of inflation we calculate the $x$-value for which
$|\eta|$ becomes unity. The loop contribution to $\eta$ is
\be
\eta_{\rm loop} = \frac{\kappa^2 \mcN}{16\pi^2} 
\( \frac{\mpl}{M}\)^2 g(x^2)
\label{eta_l}
\ee
with  
\be
g(z) =(3z+1) \ln(1+z^{-1}) +  (3z-1) \ln(1-z^{-1}). 
\ee

If $x \gg 1$, $g(x^2) \simeq -3 x^{-2} + \mcO(x^{-4})$ which implies
that $x_{\rm end} = \sqrt{3\mcN/(16\pi^2)} \kappa (\mpl/M)$.  This is
the regime where the inflaton VEV during inflation is large, and
consequently the non-renormalisable potential $V_{\rm NR}$ becomes
important.  For this reason, we will not discuss it any further.

The above expression breaks down for small enough coupling $\kappa
\lesssim 7/\sqrt{\mcN} (M/\mpl) \lesssim 10^{-2}/\sqrt{\mcN}$, (in the
last step we used $M \sim 10^{16} \GeV$, from Eq.~(\ref{M_l}) below)
and $x_{\rm end}$ approaches unity. Using $x_{\rm end} = 1$ in
Eq.~(\ref{N_Q}) we find
\be
N_Q = \frac{16 \pi^2}{k^2 \mcN} \( \frac{M}{\mpl}\)^2 
\int_1^{x_Q} \frac{1}{xf(x)} \dd x
\ee
In the limit where the factor in front of the integral ${16
\pi^2}/(k^2 \mcN) ({M}/{\mpl})^2 \gg N_Q$, the integral has to be much
less than unity which requires $x_Q \to x_{\rm end} \approx 1$.  We
conclude that we can approximate
\be
x_Q \approx x_{\rm end} \approx 1, 
\qquad 
{\rm for} \;\;\kappa \lesssim \sqrt{\frac{16 \pi^2}{N_Q \mcN}} 
\( \frac{M}{\mpl} \) = \frac{10^{-2}-10^{-3}}{\sqrt{\mcN}}.
\label{approx_x}
\ee
The inflaton VEV during inflation is always small, and NR terms can be
negligible.  Using $x_Q \approx 1$ in the expression for the
fluctuations Eq.~(\ref{dTphi}), and setting it equal to the observed
value, allows to extract $M$
\be
M_{\rm loop} 
= 2 \times 10^{-2} (\delta_C \mcN \kappa)^{1/3} \mpl,
\label{M_l}
\ee
where the subscript $loop$ is a reminder that this formula is valid in
the domain where the loop potential is the dominant term in $V'$.  The
parameter $\delta_C$ defined in Eq.~(\ref{deltaC}) gives the
normalised contribution of the inflaton sector including cosmic
strings to the CMB. If the inflaton sector including cosmic strings is
the only source of perturbations $\delta_C =1$, whereas $\delta_C \ll
1$ in the curvaton and inhomogeneous reheat scenario.  The tensor
perturbations are negligible, and in the parameter space of interest,
also the string contribution is small, $B< 10\%$.

In the loop dominated regime $M$ is a single valued function of
$\kappa$, given by Eq.~(\ref{M_l}).  This expression is valid in the
small coupling regime $\kappa < 10^{-2} - 10^{-3}/\sqrt{N}$, where it
is a good approximation to ignore the string contribution to the CMB,
and where $x_Q \approx 1$ is valid.

The CMB constraint $M_{\rm loop} < M_{\rm CMB} \approx 3\times
10^{15}/\sqrt{\epsilon(\beta)}$ discussed in section~\ref{s:CMB}
implies
\be
\kappa <
2 \times 10^{-4} \frac{1}{\delta_C \mcN \epsilon(\beta)^{3/2}}.
\label{range}
\ee
Taking into account $\epsilon(\beta)$ --- the correction to the
mass-per-unit-length away from the Bogomolny bound --- weakens the
bounds on the coupling compared to an analysis in which this
correction is ignored, as was done in Ref.~\cite{Rocher}.  This
correction changes the bound by a factor $\epsilon(\beta)^{-3/2} \sim
10$ to $\kappa \lesssim 10^{-3}$, where we have used that $\epsilon
\sim 1/5$ for $\kappa \sim 10^{-3}$.  Note, however, that the analytic
approximation breaks down in this limit, and one has to go to
numerical calculations for a more precise bound.

$V_{\rm loop}$ is always present. At small coupling it decreases
rapidly ($\propto \kappa^4$) and $V_A,V_m,V_{\rm NR}$ can become
dominant.  In the large coupling regime $S \to \mpl$ and $V_{\rm NR}$
dominates.  Note that in the regimes where other contributions
dominate, the mass scale $M$ is higher than it would be in the
presence of just the loop potential, and the CMB constraints are
stronger.  The reason is that $(\delta T/T) \propto M^6/V'$, and thus
the larger $V'$ the larger $M$ is needed to obtain the observed
temperature anisotropy.

\subsection{The non-renormalisable SUGRA regime}
\label{s:VNR}

The non-renormalisable corrections dominate over the loop corrections
for density perturbations if $V'_{\rm NR} > V'_{\rm loop}$, when
\be
\frac{x^2}{f(x^2)} > \frac{\kappa^2 \mcN}{16\pi^2} 
\(\frac{\mpl}{M}\)^4.
\ee
There are two regimes where this inequality is satisfied.  The first
is when the inflaton VEV, and thus $x$, is large during inflation.
For $x\gg1$ the term $x^2/f(x^2) \to x^4$ and the l.h.s. of the above
equation is large. For very small $\kappa$ the r.h.s. becomes small,
and this is the other regime where the NR potential can become
important.

First consider the large coupling regime where $x\gg 1$. The
contribution of $V_{NR}$ to $\eta$ is $\eta_{NR}= 3 (M/\mpl)^2 x^2$.
The slow roll parameter exceeds unity for large $x$ and drops below
one when $x = 1/\sqrt{3} (\mpl/M)$. Thus inflation can only happen for
$\sigma = \sqrt{2} M x < (2/3)^{1/2} m_p$.  At still lower $x$, the
loop potential starts dominating $\eta$; inflation ends when
$\eta_{\rm loop}$ given by Eq.~(\ref{eta_l}) becomes unity.  Hence
\be
0.2 \mpl \sqrt{\mcN} \kappa = \sigma_{\rm end} < \sigma_Q 
< 8 \times 10^{-2}  \mpl
\label{bound_xq}
\ee 
where for the upper bound we used that $|\eta| < 10^{-2}$ when
observable scales leave the horizon, to assure scale invariance, as
given by WMAP data \cite{WMAP}. Both inequalities together give the
upper upper bound $\kappa < {0.5}/{\mcN}$ obtained in the limit
$x_Q\to x_{\rm end}$.  One should realize however, that 60 e-folds
should pass when the inflaton rolls from $x_Q$ to $x_{\rm end}$, and
they cannot be taken arbitrarily close together.

An analytical approximation in the large coupling regime is hard
because of a non-trivial expression for $\sigma_Q$, and because the
string contribution to the CMB becomes important.  However, in the
range where non-renormalisable corrections dominate, the scale $M$ is
higher than $M_{\rm loop}$ given in Eq.~(\ref{M_l}), which is excluded
by CMB measurements, i.e., by the requirement $M < M_{\rm CMB}$. The
spectral index is
\be 
n - 1 \simeq  2 \eta 
= - {3 \sigma_Q^2\over \mpl^2} 
+  {3\kappa^2 \mcN \mpl^2 \over 4 \pi^2 \sigma_Q^2}; 
\label{nNR}
\ee
it goes from negative to positive as the non-renormalisable
contribution comes to dominate. The running of the spectral index is
small \cite{run} : $\dd n/\dd\ln \kappa \sim -10^{-3}$.

In the small coupling regime where $\kappa < 3 /\sqrt{\mcN}
(M/\mpl)^2$ the NR potential dominates over the loop potential.  As
before, for small $\kappa$ it is a good approximation to take $x_Q
\approx x_{\rm end} \approx 1$ (see Eq.~(\ref{approx_x})).  Then
Eq.~(\ref{dTphi}) gives
\be
M_{\rm NR} = 
3 \times 10^{14} \GeV \( \frac{\kappa}{10^{-7}} \) W^{-1/2}
\label{M_NR}
\ee

\begin{figure}
\begin{center}
\leavevmode\epsfysize=7cm \epsfbox{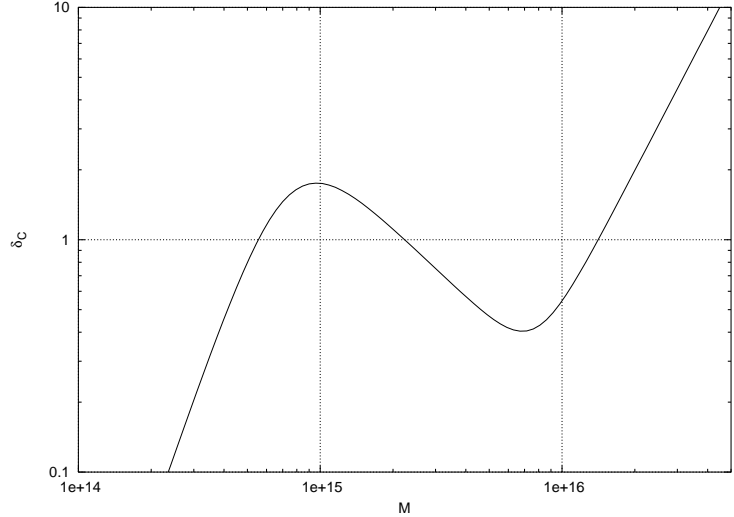}
\caption{$\delta_C$ vs. $M$ for $\kappa =10^{-6}$ and $\mcN =1$.}
\label{F:dT}
\end{center}
\end{figure}

It is possible to have for one value of $\kappa$ several solutions for
$M$, which all produce the correct density perturbations.  The reason
is that for fixed $\kappa$, by varying $M$, different contributions
dominate the density perturbations.  At small $M$ the loop potential
dominates and $(\delta T/T) \propto M^3$, at larger $M$ the NR
contribution becomes important and $(\delta T/T) \propto M^{-1}$, and
at still larger $M$ the main contribution comes from cosmic strings
and $(\delta T/T) \propto M^2$.  An example is given in
Fig.~\ref{F:dT}, where we have plotted the $\delta_C$ --- the
temperature fluctuation normalized by the COBE value as defined in
Eq.~(\ref{deltaC}) --- as a function of $M$ for $\kappa = 10^{-6}$.
The maximum $M_1$ corresponds to the mass scale where $V'_{\rm loop} =
V'_{\rm NR}$, whereas the minimum $M_2$ corresponds to the scale where
$(\delta T/T)_{\rm inf} = (\delta T/T)_{\rm cs}$.  Now
\bea
M_1 &=& 0.3 \kappa^{1/2} \mcN^{1/4} \mpl \nonumber \\
M_2 &=& 0.3 \( \frac{0.9}{y \epsilon} \)^{1/3} \kappa^{1/3} \mpl
\eea
where in the second line we have used that $\epsilon \sim 10^{-1}$ for
small $\kappa$. For simplicity we have set $\delta_C=1$, and assume
that no alternative mechanism for density perturbations such as the
curvaton scenario is at work. We will return to this point at the end
of the subsection. We define $\kappa_1$ the coupling for which
$\delta_C(M_1) =1$, $\kappa_2$ the coupling for which $\delta_C ( M_2)
=1$.
\bea 
\kappa_1 &=& 6\times 10^{-8} \mcN^{1/2}
\nonumber \\ 
\kappa_2 &=& 4\times 10^{-6} \( \frac{0.9}{y \epsilon}\)^{1/2} 
\eea
There are then 3 possibilities.
\begin{enumerate}
\item {$\delta_C(M_1)>1$ and $\delta_C(M_2)>1 \quad
\Longleftrightarrow \quad \kappa > \kappa_1,\;\kappa_2$}\\ There is
only one solution $M(\kappa)$ for which the loop potential
dominates. It is given by $M_{\rm loop}$ in Eq.~(\ref{M_l}). Note that
at still larger coupling the contribution from NR terms and strings
kick in again.
\item {$\delta_C(M_1)>1$ and $\delta_C(M_2)<1 \quad
\Longleftrightarrow \quad \kappa_2> \kappa > \kappa_1$}\\ There are
three solutions.  One at low scale where the loop potential dominates
and $M$ is given by $M_{\rm loop}$ of Eq.~(\ref{M_l}), a middle one
where NR terms dominate and $M$ is given by $M_{\rm NR}$ of
Eq.~(\ref{M_NR}), and one at high scale which is dominated by string
contributions.  The latter solution is excluded by CMB data.
\item {$\delta_C(M_1)<1$ and $\delta_C(M_2)<1\quad \Longleftrightarrow
\quad \kappa < \kappa_1,\;\kappa_2$}\\ There is one solution,
dominated by the string contribution. This solution is excluded by CMB
data.
\end{enumerate}

Note that $\kappa > \kappa_1,\kappa_2$ is a lower bound on the
coupling to be consistent with CMB data.  Thus in the presence of NR
terms, the coupling cannot be arbitrarily small.  The minimum mass
scale is $M_{\rm min} = M_{\rm loop}(\kappa_1) =2 \times 10^{14}
\sqrt{\mcN} $.

\begin{figure}
\begin{center}
\leavevmode\epsfysize=8cm \epsfbox{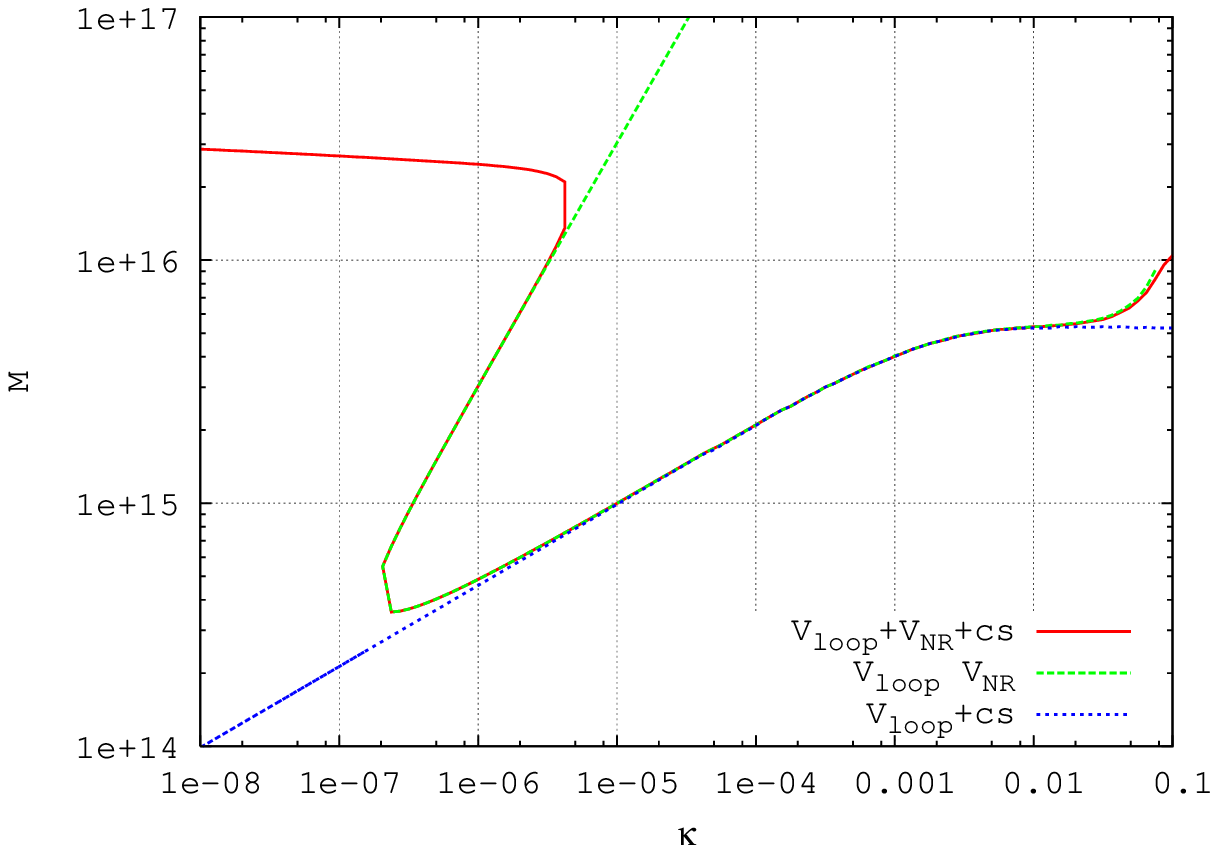}
\caption{$M$ vs. $\kappa$ for $\mcN =1$.}
\label{F:ns}
\end{center}
\end{figure}

The above discussion is illustrated by the numerical results shown in
Fig.~\ref{F:ns}.  Matching the quadrupole to the observed value gives
$M$ as a function of $\kappa$. The result with $V_{\rm loop}$, $V_{\rm
NR}$ and the string contribution all turned on is given by the solid
line.  Also plotted are the result with $V_{\rm NR}$ turned off
(dashed line) and with the string contribution turned off (dotted
line).  Consider the full solution, the solid line.  At large coupling
both the NR and string contribution are important.  There is no
solution for arbitrarily large $\kappa$, in agreement with the
discussion below Eq.~(\ref{bound_xq}). Going to smaller coupling, the
loop potential starts to dominate and $x_Q \to 1$: the solution is a
straight line as given by $M_{\rm loop}$ in Eq.~(\ref{M_l}).  The
solution $M_{\rm loop}$ extends at low coupling until $\kappa =
\kappa_1 \sim 5 \times 10^{-7}$.  The second branch, where NR terms
dominate, is between $5 \times 10^{-7} \sim \kappa_1 < \kappa
<\kappa_2 \sim 5 \times 10^{-6}$.  The upper branch is string
dominated and excluded by CMB data.  In the presence of NR terms
agreement with the CMB requires $\kappa > \kappa_1$ and $M_{\rm min}
\sim 4 \times 10^{14}< M_{CMB}$.

Let us now briefly return to the possibility that alternative
mechanisms for density perturbations are at work and $\delta_C \ll
1$. As can be seen from Fig.~\ref{F:dT}, for small $\delta_C$ there is
only one solution which is dominated by the loop contribution, and
thus given by $M_{\rm loop}$ in Eq.~(\ref{M_l}).

\subsection{The Hubble regime}

Consider the Hubble induced mass term in $V_{\rm m}$. This term
dominates the density perturbations for $V'_{\rm m} > V'_{\rm loop}$,
or
\be
\kappa < \frac{4 \pi |a|}{\sqrt{\mcN f(x^2)}} \( \frac{M}{\mpl} \) 
\approx
10^{-1} \frac{|a|^{3/2} \delta_C^{1/2}}{\mcN^{1/4}}
\ee
In the second equality we took $M=M_{\rm loop}$ and $f(x^2) \sim 1.4$,
which is valid for $\kappa < 10^{-2}-10^{-3}/\sqrt{N}$.  The Hubble
induced term dominates for small enough $\kappa$.  If domination
happens for $\kappa < \kappa_1 \sim 5 \times 10^{-6}$, the region
already excluded by CMB date due to NR terms, the Hubble induced mass
plays no r\^ole.  This is for $|a| < a_0 \sim 10^{-3}
(\mcN/\delta_C^2)^{1/6}$.

Consider then $|a| > a_0$; the Hubble induced mass becomes important
before NR terms kick in.  The contribution of the Hubble induced mass
to $\eta$ is constant $\eta_m = |a|^2$; scale invariance requires $|a|
< 0.1$.  The end of inflation is determined by the loop contribution,
and $x_{\rm end} \approx 1$. Further $x_Q \approx 1$ for small
coupling.  Plugging this into Eq.~(\ref{dTphi}) gives
\be
M_{\rm m} = 8 \times 10^{-4} \mpl \frac{|a|^2 \delta_C}{\kappa} .
\label{M_m}
\ee
In the $\kappa$-region where $V'_{\rm m}$ dominates $M$ is a single
valued function of $\kappa$.  Since $M_{\rm m}$ grows with decreasing
coupling, whereas $M_{\rm loop}$ decreases with decreasing coupling,
there is a minimum scale $M$, obtained for $V'_{\rm loop} = V'_{\rm
m}$, or equivalently, for $M_{\rm loop} = M_{\rm m}$:
\be
M_{\rm min} = 2 \times 10^{16} \sqrt{|a|\delta_C}\mcN^{1/4}.
\ee
Agreement with CMB data requires $M_{\rm min} > M_{\rm CMB}\sim
3\times 10^{15}\GeV/\sqrt{\epsilon(\beta)}$, or
\be
|a| < \frac{2 \times 10^{-2}}{\sqrt{\mcN} \delta_C} 
\frac{1}{\epsilon(\beta)}.
\ee
This gives a stronger bound on $|a|$ than the requirement of scale
invariance.

\subsection{The $A$-regime}

The $A$-term breaks the discrete symmetry $S \leftrightarrow -S$.  For
$A <0$ there is a minimum for the potential at $S_0 \neq 0$.  If $S_0
> S_c$ the inflaton gets trapped in the false vacuum leading to
eternal inflation. Assume $V = V_0 + V_{\rm loop} + V_{\rm A}$, and
all other terms negligible small. Then this happens for $x_0 > x_c$
with $x_0 \sim (\mcN \kappa^3 M)/(8\pi^2 A m_{3/2})$ and $x_c =1$,
which gives
\be
M \gtrsim \frac {8\pi^2 A}{\mcN} \frac{m_{3/2}}{\kappa^3}
\sim \frac{10^{16} \GeV}{\mcN}  \(\frac{m_{3/2}}{10^2 \GeV} \)
\( \frac{10^{-4}}{\kappa} \)^3
\ee
Hence for a negative $A$-term, this is a problem for large
$\kappa$/small gravitino mass.  

No new minimum arises for positive $A$-term, which is the case we
discuss now.  These results equally apply to a negative $A$-term
provided $x_0 \ll 1$.  The $A$-term dominates the density
perturbations if $V'_A>V'_l$, or
\be
\kappa < 4 \( \frac{ m_{3/2} A}{ \mcN M} \)^{1/3} \sim 
\frac{4 \times 10^{-5} }{\mcN^{2/5} \delta_C^{1/10}}
\( \frac{m_{3/2} A}{\GeV} \)^{3/10}
\ee
where in the second step we used $M \sim M_{\rm loop}$.  A-term
domination at small coupling happens before NR terms become important,
unless the gravitino mass is small $m_{3/2} < 5\times 10^{-4} \GeV
(\mcN/A)$. Consider then $m_{3/2}$ large enough for the linear term to
play a r\^ole. The A-term does not contribute to $\eta$, and $x_{\rm
end} \approx 1$ determined by the loop contribution.  In addition $x_Q
\approx 1$ for small coupling.  Using Eq.~(\ref{dTphi}), the observed
perturbations are obtained for
\be
M_A = \frac{4 \times 10^{13} \delta_C^{1/4}}{\sqrt{k}} 
\( \frac{m_{3/2} A}{10^3 \GeV}\)^{1/4} 
\label{M_A}
\ee
When the linear term dominates, $M$ increases as a function of
decreasing coupling.  This gives a lower bound on the mass scale $M$,
obtained for $V'_A = V'_l$, which is
\be
M_{\rm min} = 2 \times 10^{15} \GeV \mcN^{1/5} \delta_C^{1/20} 
\(\frac{m_{3/2} A}{10^3\GeV} \)^{1/10}
\ee
This is only consistent with the CMB bound, i.e., with the requirement
$M_{\rm min} > M_{\rm CMB}$, for $\mcN =1$ and $m_{3/2} \lesssim 10^3
\GeV$.  Note that the dependence on the gravitino mass $m_{3/2}$ is
weak, and very small $m_{3/2}$ is needed to lower this minimum scale
considerably.  This favors gauge mediated SUSY breaking, or a scenario
in which hidden sector SUSY breaking happens after inflation.

\section{Numerical results}
\label{s:numerical}

We have also solved the equations pertaining the density perturbations
Eqs.~(\ref{y},\ref{N_Q},\ref{slow_roll},\ref{dTphi},\ref{dTtot}).
For $\epsilon(\beta)$ we use the second expression in
Eq.~(\ref{epsilon}).  This introduces a factor 4 error in the limit
$\beta = (\kappa/g)^2 \to 1$, where the BPS limit $\epsilon(1) =1$
should be approached.  Further we take $y=3$, where $y$ is the
parameter which parameterises the string contribution as given in
Eq.~(\ref{y}), and reconnection probability $p=1$.  Setting the
quadrupole Eq.~(\ref{dTtot}) equal to the observed value given in
Eq~(\ref{cobe}) gives the symmetry breaking scale $M$ as a function of
$\kappa$.  All other quantities such as the spectral index and the
inflaton VEV when observable scales leave the horizon can also be
computed. In this section we discuss the results.

Fig.~\ref{F:dim} shows $M$ vs. $\kappa$ for $\mcN =1,16,126$. The
potential includes the loop potential and the NR terms; all other
terms are turned off.  Also shown are the bounds on the scale $M$;
from top to bottom these are the 10\%-bound, the
Kaiser-Stebbins-bound, and the pulsar bound. The 10\% bound and
especially the pulsar bound has the large theoretical uncertainties.
At large and small coupling cosmic strings dominate the density
perturbations, or equivalently $M(\kappa)$ exceeds the various bounds.
This is the region excluded by experiments. At intermediate coupling,
and for small enough $\mcN$, there are then two branches compatible
with CMB data, the branch where the loop potential dominates, and a
second branch at small coupling where the NR terms dominate.  This in
good agreement with the discussion in section~\ref{s:VNR}.

The $\kappa$ range compatible with all bounds is $10^{-6}\lesssim
\kappa \lesssim 10^{-3}/\mcN$.  The upper bound lies in the
$\kappa$-range where $V'_{\rm loop}$ dominates, which is why it is
$\mcN$ dependent.  On the other hand, the lower bound is determined by
$V'_{\rm NR}$ which is independent of $\mcN$.  If we drop the pulsar
bound, the parameter range is extended at large coupling to
$10^{-7}/\mcN \kappa \lesssim 10^{-2}/\mcN$. Only a small window
remains for $\mcN = 126$, which is clearly disfavored.

In Fig.~\ref{F:index} we show the spectral index for the parameters of
Fig.~\ref{F:dim}. The spectral index is less than one, except at large
coupling when NR terms start to dominate the second derivative, see
Eq.~(\ref{nNR}).  The coupling for which the spectral index starts to
diverge corresponds to the upper bound on $\kappa$ for which a
solution exists (compare Figs.~\ref{F:dim},~\ref{F:index}).  This is
in agreement with the discussion below Eq.~(\ref{bound_xq}), where it
should be noted that the potential is steep for large $\eta$ and thus
$x_{\rm end}$ and $x_Q$ are well separated.  The spectrum is
indistinguishable from scale invariance and gives no new constraints
on the available parameter space.

\begin{figure}
\begin{center}
\leavevmode\epsfysize=8cm \epsfbox{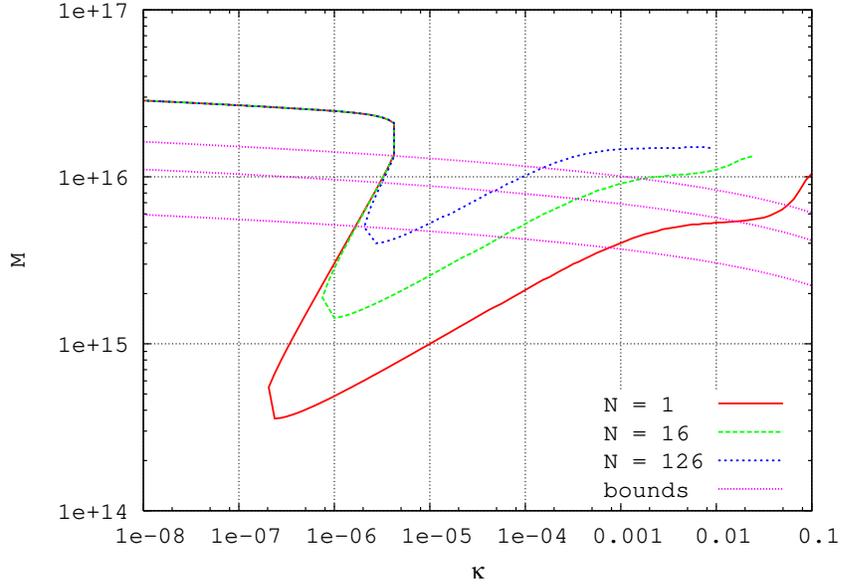}
\caption{$M$ vs. $\kappa$ for $\mcN =1,16,126$. Further shown are,
from top to bottom, the 10\%-bound, the KS-bound and the pulsar
bound.}
\label{F:dim}
\end{center}
\end{figure}

\begin{figure}
\begin{center}
\leavevmode\epsfysize=8cm \epsfbox{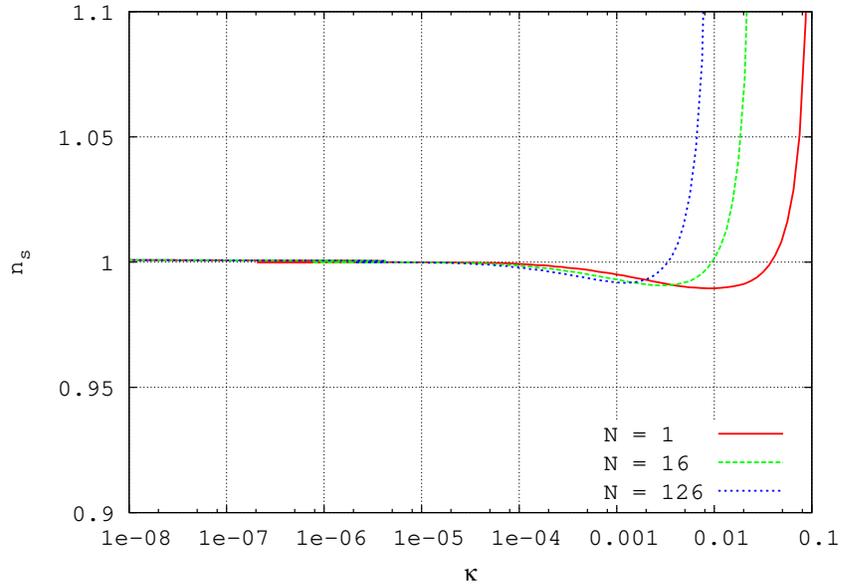}
\caption{Spectral index $n$ vs. $\kappa$ for $\mcN =1,16,126$}
\label{F:index}
\end{center}
\end{figure}

The string contribution to the quadrupole is parametrized by the
parameter $y$.  The value for $y$ found in the literature ranges from
$y=3-12$.  Fig~\ref{F:B} shows how this affects the 10\%-bound.
Plotted is the string contribution to the quadrupole $B$ defined in
Eq.~(\ref{B}) as a function of $\kappa$ for $\mcN =1$ and different
values of $y$.  The 10\% bound corresponds to $B = 0.1$.  For $y=3$
the bound on $\kappa$ corresponds to that found from the $M(\kappa)$
plot in Fig.~\ref{F:dim}, as it should. For $y=9,12$ the 10\%-bound is
stronger, and the upper bound on $\kappa$ is decreased by about a
factor 10.  We want to stress that in contrast to the 10\% bound,
the KS and pulsar bound are rather insensitive to $y$.  And thus the
$\kappa$-range compatible with the KS and pulsar bound is practically
the same for the different $y$-values.

\begin{figure}
\begin{center}
\leavevmode\epsfysize=8cm \epsfbox{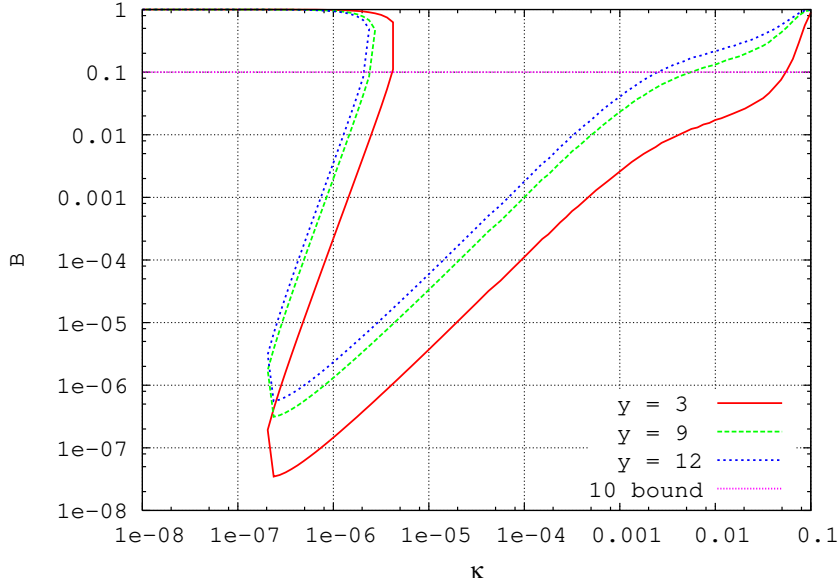}
\caption{$B$ vs. $\kappa$ for $\mcN=1$ and $y =3,9,12$. $B=0.1$
corresponds to the 10\%-bound.}
\label{F:B}
\end{center}
\end{figure}

Fig~\ref{F:m} shows the results when the Hubble induced mass term is
included, the different lines corresponds to different values of $|a|
=10^{-1},10^{-2},10^{-3}$; in all cases $\mcN =1$. $M(k)$ increases
with increasing $|a|$. Indeed too large $|a| = 10^{-1}$ is excluded by
the data, whereas $|a|\sim 10^{-2}$ decreases the upper bound
considerably, to $\kappa \lesssim 10^{-5}$.  For $|a| \lesssim
10^{-3}$, the available parameter space is unaltered.  $V'_m$ is
independent of $\mcN$, and we can get a good handle on how it affects
parameter space for different $\mcN$ by comparing Figs.~\ref{F:dim}
and \ref{F:m}. For $\mcN =16,126$ it follows that $|a| \gtrsim
10^{-2}$ is excluded, whereas for example $|a| \sim 10^{-3}$ reduces
the upper bound to $\kappa \lesssim 10^{-5}, 10^{-6}$.

\begin{figure}
\begin{center}
\leavevmode\epsfysize=8cm \epsfbox{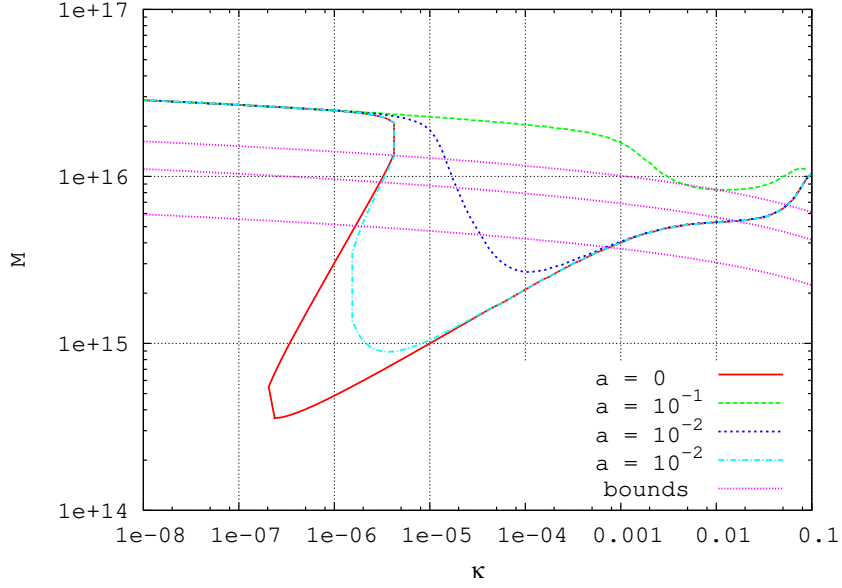}
\caption{$M$ vs. $\kappa$ with Hubble mass term included. The plots
are for $a =10^{-1},10^{-2},10^{-3}$. Also shown are, from top to
bottom, the 10\% bound, the KS-bound and the pulsar bound.}
\label{F:m}
\end{center}
\end{figure}

Fig.~\ref{F:ms} shows the effect of including the linear A-term for
gravitino masses $m_{3/2} = 10^3,10^2,10^0,10^{-2}\GeV$ and $\mcN
=1$. A gravitino mass $m_{3/2} \gtrsim 10^2 \GeV$ as obtained in
gravity mediated SUSY breaking shrinks parameter space considerably:
for $\mcN = 1$ the allowed range of $\kappa$ is $10^{-5}-10^{-4}
\lesssim \kappa \lesssim 10^{-3}-10^{-2}$, whereas no parameter space
is left for $\mcN = 16,126$.

\begin{figure}
\begin{center}
\leavevmode\epsfysize=8cm \epsfbox{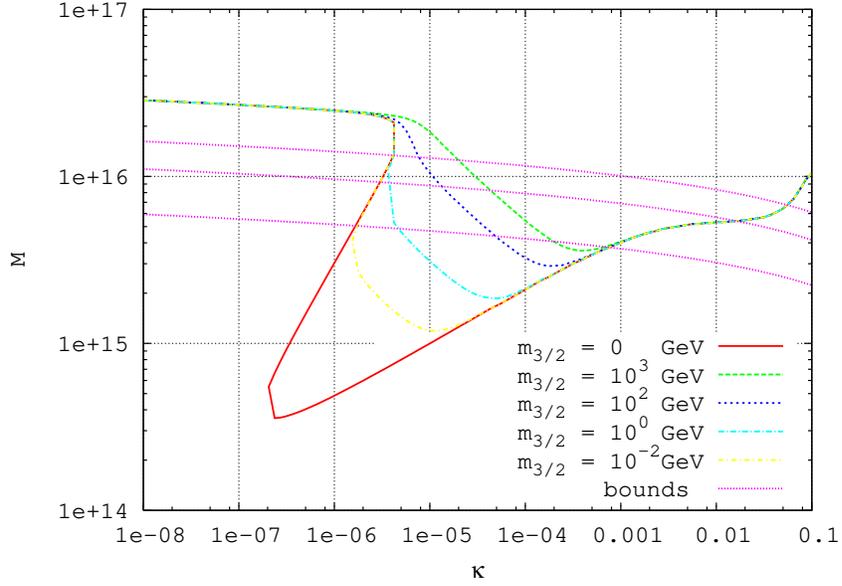}
\caption{$M$ vs. $\kappa$ with A term included. The plots are $m_s
=10^{3},10^{2},10^{0},10^{-1}$. Also shown are, from top to bottom,
the 10\% bound, the KS-bound and the pulsar bound.}
\label{F:ms}
\end{center}
\end{figure}

\begin{figure}
\begin{center}
\leavevmode\epsfysize=8cm \epsfbox{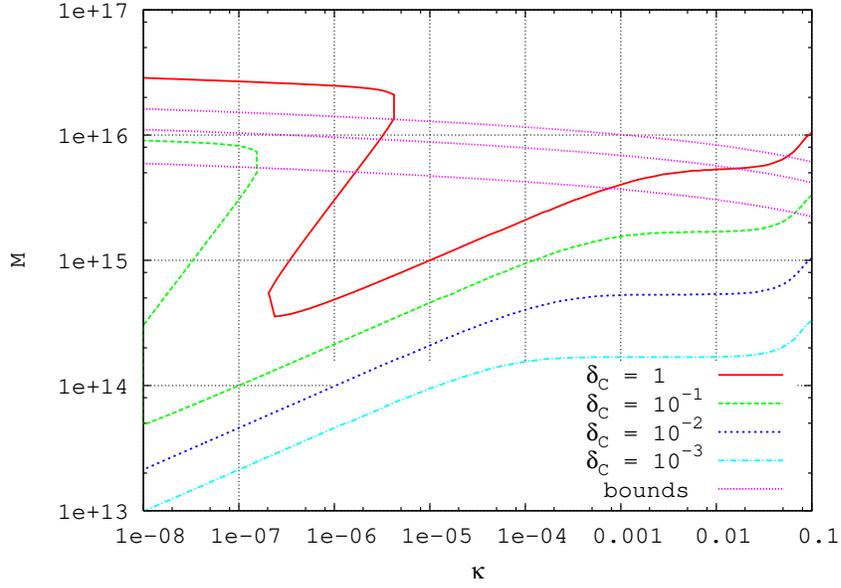}
\caption{$M$ vs. $\kappa$ for $\delta_C =
10^{-1},10^{-2},10^{-3}$. Also shown are, from top to bottom, the
10\% bound, the KS-bound and the pulsar bound.}
\label{F:curv}
\end{center}
\end{figure}

\begin{figure}
\begin{center}
\leavevmode\epsfysize=8cm \epsfbox{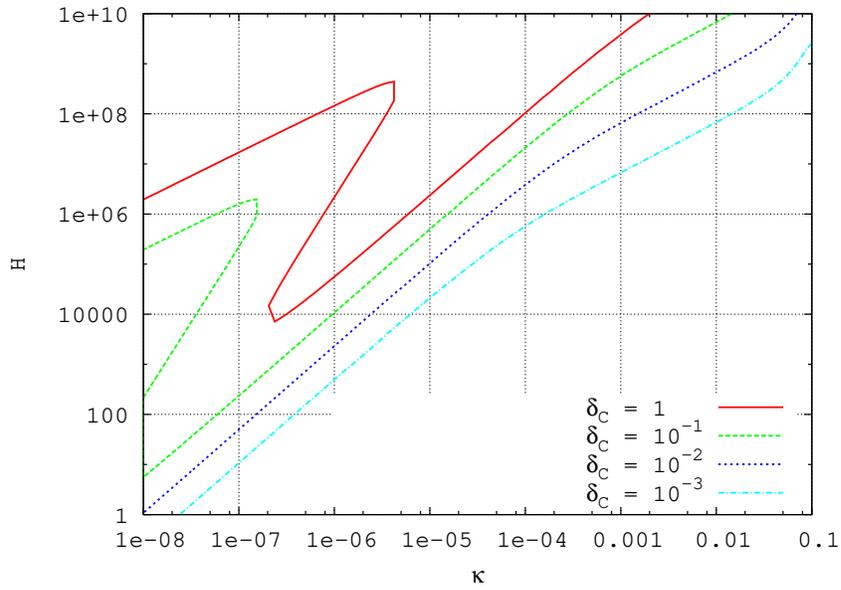}
\caption{$H$ vs. $\kappa$ for $\delta_C = 10^{-1},10^{-2},10^{-3}$.}
\label{F:H}
\end{center}
\end{figure}

Finally we consider the possibility that an other scalar field than
the inflaton gives the main contribution to the density perturbations,
as occurs in the curvaton or inhomogeneous reheat scenario.  If
Fig.~\ref{F:curv} we plot $\kappa$ vs. $M$ for $\delta_C=
10^{-1},10^{-2},10^{-3}$.  Already when the combined contribution of
the inflaton and cosmic strings to the quadrupole is reduced to 10\%
of the COBE value (and thus the other 90\% are provided by some other
scalar field), are all constraints avoided.  Invoking an alternative
scalar field to explain the density perturbations is thus a good way
to avoid all constraints on the string scale. Note however, that the
curvaton scenario in its simplest form can only work for Hubble
constants $H > 10^{7}\GeV$.  As can be seen from Fig.~\ref{F:H}, where
we plotted $H$ vs. $\kappa$ for the same parameters as in
Fig.~\ref{F:curv}, this implies $\kappa > 10^{-3}-10^{-4}$.  No such
constraint exists for the inhomogeneous reheat scenario.

\subsection{Brane $F$-inflation}

In the plots discussed above we have used the GUT value for the gauge
coupling $g$ and varied the trilinear coupling $\kappa$. In brane
models the gauge coupling may differ by few orders of magnitude from
the GUT value and the coupling constant is given in terms of $g$.
This does not change anything for the inflation contribution (which is
now given as function of $g$ instead of $\kappa$) but only for the
strings.

Here we discuss as an example $F$-term inflation that arises as a
certain limit of $P$-term brane inflation models \cite{Pterm}.
$P$-term inflationary models in $N=1$ supergravity interpolate between
$F$- and $D$-term models.  The choice of a particular model is
determined by the VEV of the auxiliary triplet field of P-inflation,
which depends on the fluxes on the branes \cite{Pterm}.

The matter content of $P$-inflation is an $N=2$ charged hypermutiplet
and a $U(1)$ gauge multiplet. These contain in addition to the gauge
bosons a pair of complex conjugate fields which can be identified with
our $\phi_+$ and $\phi_-$ fields and a neutral singlet which we denote
by $S$. The $N=1$ superpotential is given by
\be
W = \sqrt{2} g S (\phi_+ \phi_- - M^2).
\ee
Hence, we recover the superpotential given in Eq.~(\ref{W}) with
$\kappa = \sqrt{2} g$. The strings that form at the end of $F$-term
$P$-inflation have $m_A = m_\phi$, and satisfy the Bogomolny bound.
Their tension is $\mu = 2 \pi M^2$, i.e., $\epsilon(\beta)$ in
Eq.~({\ref{mu}) equals unity. This in turn modifies the parameter
range allowed by the data. In Fig.~\ref{F:Pterm} we plot $M$ versus
$\kappa$ for $\epsilon(\beta)=1$. The 10\%-bound, the KS-bound, and
the pulsar bound on $M$ are proportional to $\epsilon(\beta)^{-1/2}$
(see Eqs.~(\ref{Pogosian},~\ref{smoot},~\ref{pulsar}), and are more
stringent than for GUT F-term inflation.  The KS-bound gives $10^{-6}
\lesssim g \lesssim 10^{-4}$ which is much below the expected range
for $g$.  \footnote{The 10\% bounds implies $g < 10^{-3}$ which
differs from result $g <10^{-4}$ quoted in \cite{Rocher}.  This
difference can be traced back to the different $y$ values used: $y=3$
in our case and $y=9$ in \cite{Rocher}}

\begin{figure}
\leavevmode\epsfysize=8cm \epsfbox{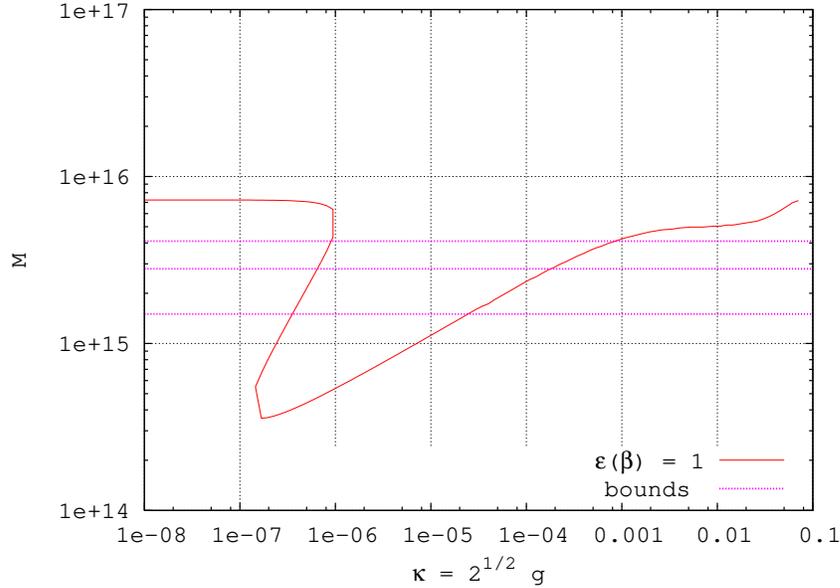}
\caption{$M$ vs. $\kappa$ for $\mcN =1$ and $\epsilon(\beta)=1$
corresponding to $F$-term $P$-inflation. Also shown are, from top to
bottom, the 10\% bound, the KS-bound and the pulsar bound.}
\label{F:Pterm}
\end{figure}

\section{Warm inflation}
\label{s:warm}

If the inflaton, or a field coupling to the inflaton, can decay during
inflation, it has a propagator of the Breit-Wigner form with an
imaginary part related to the decay rate $\Gamma$.  Upon calculating
the one-loop correction to the inflaton potential, these contributions
related to the decay rate lead to dissipative effects.  In the
adiabatic-Markovian limit, satisfied for $\dot{\phi}/\phi < H <
\Gamma$, this can be described by an effective friction term
$\Upsilon$ in the inflaton equation of motion~\cite{berera}
\be \ddot{S} + (3H + \Upsilon)\dot{S} + \frac{\partial V}{\partial S}
= 0.  \ee
In hybrid inflation inflaton decay is kinematically forbidden.
However, the heaviest Higgs field $\chi_+ = (\phi_+ +
\phi_-)/\sqrt{2}$ can decay into its fermionic superpartner
$\tilde{\chi}_+$ and an inflatino.  Since the masses of $\chi_+$ and
$\tilde{\chi}_+$ are close together, this decay is phase space
suppressed. As shown in Ref.~\cite{mar} its effects can be neglected
during inflation.  Dissipation can only be important if the inflaton
sector couples to light particles.  We consider the coupling
\be
W = \lambda \phi_+ NN.
\label{RHN}
\ee
If hybrid inflation is embedded in a grand unified theory, and the
$U(1)$ broken at the end of inflation corresponds to baryon$-$lepton
number, the above term naturally arises as a mass term for the right
handed neutrino $\tilde{N}$.

Physically, what happens is that during inflation the slowly changing
inflaton field can excite the Higgs field $\phi_+$, which can decay
into massless $N$ fields.  Through this channel, the inflaton sector
dissipates radiation: $ \dot{\rho}_\gamma + 4 H \rho_\gamma = \Upsilon
\dot{S}^2 $.

One can distinguish three regimes, depending on the strength of the
effective damping term $\Upsilon$ and on the temperature $T \sim
\rho_\gamma^{1/4}$ of the radiation bath.
\begin{enumerate}
\item{$\Upsilon,\,T < H$.}  This is the regime of cold inflation where
dissipation is negligible small.  The Hubble parameter sets the
friction term in the inflaton equation of motion, as well as the
scale of inflaton perturbations $\delta S^2 \propto H^2$.  
\item{$H <T<\Upsilon$.} Warm inflation in the weak dissipative
regime.  The friction term is dominated by the Hubble constant, i.e.,
by the expansion of the universe, but the perturbations are dominated
by thermal effects and $\delta S^2 \propto {HT}$.
\item{$H<\Upsilon$.} Warm inflation in the strong dissipative regime.
The damping of the inflaton field is dominated by $\Upsilon$, i.e.,
by dissipative effects. Moreover, fluctuations are thermal with $\delta
S^2 \propto \sqrt{H\Upsilon}T$.
\end{enumerate}

Warm inflation, which occurs for sufficiently large couplings
$\kappa,\lambda$, lowers the scale $M$ for a given $\kappa$ compared
to cold inflation.  The reason is that the fluctuations $\delta S$ are
larger by a factor $\sqrt{T/H} (1+ (\Upsilon/H)^{1/4})$, so that a
smaller $M$ is required to obtain the right temperature anisotropy.
In the strong dissipative regime there is the additional effect that
the inflaton is stronger damped as compared to cold inflation, and the
inflaton value when observable scales leave the horizon, $S_Q$, is
lowered.

We will list here the important formulas governing the density
perturbations; more details can be found in \cite{mar,berera}.  The
decay rate for the process $\phi_+ \to NN$ is $\Gamma = \lambda^2
m_+/(16\pi) $, with $m_+^2 = \kappa^2(|S|^2 + M^2)$ and $\lambda$ the
coupling in Eq.~(\ref{RHN}). We can define the ratio $r \equiv
\Upsilon/(3H)$ which is given by
\be
r(x) = \frac{\kappa^2}{128\sqrt{3}\pi} \(\frac{\lambda^2}{16\pi}\) 
\frac{x^2}{\sqrt{1+x^2}} \frac{\mpl}{M}
\ee
The dissipative effects parametrized by $r(x_Q)$ are maximized in the
limit $\kappa,\lambda \to 1$ large, as this maximizes the decay rate.
The formulas for the density parameters then generalize as follows.
The number of e-folds is
\be 
N_Q = \int_{\sigma_{\rm end}}^{\sigma_Q} 
\frac{1}{\mpl^2} \frac{V}{V'}(1+r) \dd \sigma.
\label{w:N_Q}
\ee
The slow roll parameters change to $ \epsilon_\Upsilon =
\epsilon/(1+r)^2$ , $ \eta_\Upsilon = \eta/(1+r)^2 $, with
$\epsilon,\eta$ given in Eq.~(\ref{slow_roll}).  In addition a third
slow roll parameter can be defined $ \bar{\epsilon}_\Upsilon =
\beta_\Upsilon r /(1+r)^3$ with
\be
\beta_\Upsilon = \mpl^2 \frac{V'}{V} \frac{\Upsilon'}{\Upsilon}.
\ee
Inflation ends when one of the slow roll parameters exceeds unity, or
when the energy inflaton decay products $\rho_\gamma$ becomes larger
than $V_0$, with
\be
\frac{\rho_\gamma}{H^4} = \frac{9}{2} 
\frac{r \epsilon}{(1+r)^2\kappa^2} 
\(\frac{\mpl}{ M}\)^4
\ee
One can define a corresponding temperature $T \approx
\rho_\gamma^{1/4}$.  The density perturbations generalize to
\be
\( \frac{\delta T }{T} \)_\phi = 
\frac{1}{12\sqrt{5}\pi \mpl^3}\frac{V^{3/2}}{V'} 
\(1 + \sqrt{\frac{T}{H}}\) \(1+ \(\frac{\pi \Upsilon}{4H} \)^{1/4} \)
\label{w:dTphi}
\ee
In the limit $r \to 0$ (and thus also $\Upsilon \to 0$, $T \to 0$) all
above formulas reduce to those of standard cold inflation, where
dissipative effects are negligible small.  Finally, we give the
spectral index in the various regimes
\be
n_s - 1 =
\left\{
\begin{array}{lll}
& -6 \epsilon + 2 \eta,  
&\qquad {\rm for} \; \Upsilon,T < H \\
& -\frac{17}{4} \epsilon + \frac32 \eta - \frac14 \beta_\Upsilon,  
&\qquad  {\rm for} \; \Upsilon < H < T \\
& (-\frac{9}{4} \epsilon + \frac32 \eta - \frac94 \beta_\Upsilon) 
(1+r)^{-1},  
&\qquad  {\rm for} \; H < \Upsilon,T 
\end{array}
\right.
\label{n_warm}
\ee

In Fig.~\ref{F:warm} we show the effects of dissipation for different
values of $\lambda =1,0.1,0.01,0$ and $\mcN =1$. As expected
dissipation is only important for large $\kappa$ and $\lambda$, and
the scale of inflation is lowered.  For $\lambda \gtrsim 0.1$ all
bounds are evaded. For smaller coupling the importance of the
dissipative effects diminishes, e.g. for $\lambda \gtrsim 0.01$ the
bound is only slightly improved.  The spectral index for the same
parameters is shown in Fig.~\ref{F:warmn}. The spectrum is scale
invariant over the whole $\kappa$ range consistent with CMB data. For
$\lambda \sim 1$ there is a discontinuity in the spectral index, which
corresponds to the transition from the weak to the strong dissipative
regime; it is merely an artifact of the approximation used in
Eq.~(\ref{n_warm}).

Fig.~\ref{F:warm16} shows the mass scale $M$ as a function of
$\kappa$, now for $\lambda = 1,0.5,0.1,0$ and $\mcN=16$.  The effect
of dissipation is smaller than for $\mcN =1$.  Only for larger, order
one, couplings are all bounds evaded. Already for $\lambda =0.1$ there
is no enlargement of parameter space.

Lastly, we show in Fig.~\ref{F:warmp} the effects of dissipation for
the Bogomolny strings arising in $F$-term $P$-inflation, for which
$\epsilon(\beta) =1$.  Plotted is $M$ vs. $\kappa$ for $\lambda =
1,0.1,0.01,0$.  Just as for GUT strings with $\mcN=1$ are the bounds 
considerably improved for $\lambda < 0.1$.

\begin{figure}
\begin{center}
\leavevmode\epsfysize=8cm \epsfbox{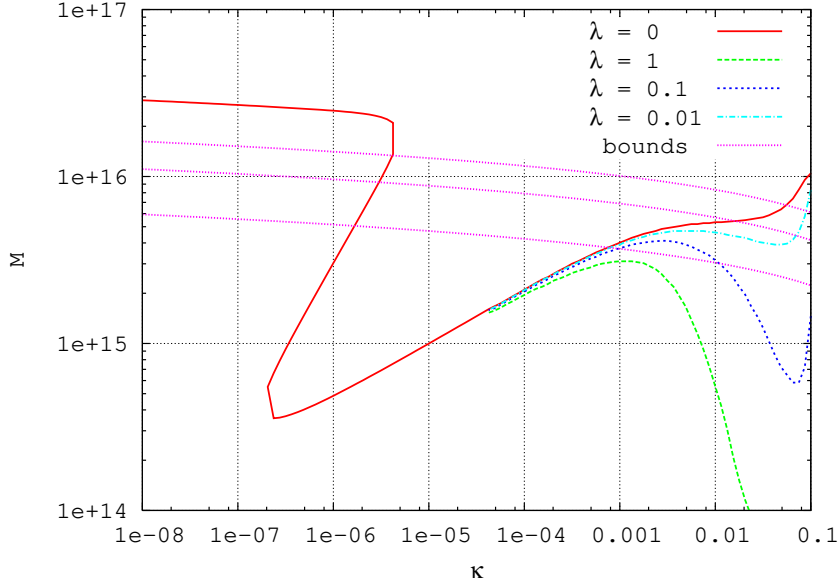}
\caption{$M$ vs. $\kappa$ for $\lambda =1,0.1,0.01,0$ and $\mcN
=1$. Also shown are, from top to bottom, the non-Gaussianity bound,
the 10\% bound, and the pulsar bound.}
\label{F:warm}
\end{center}
\end{figure}

\begin{figure}
\begin{center}
\leavevmode\epsfysize=8cm \epsfbox{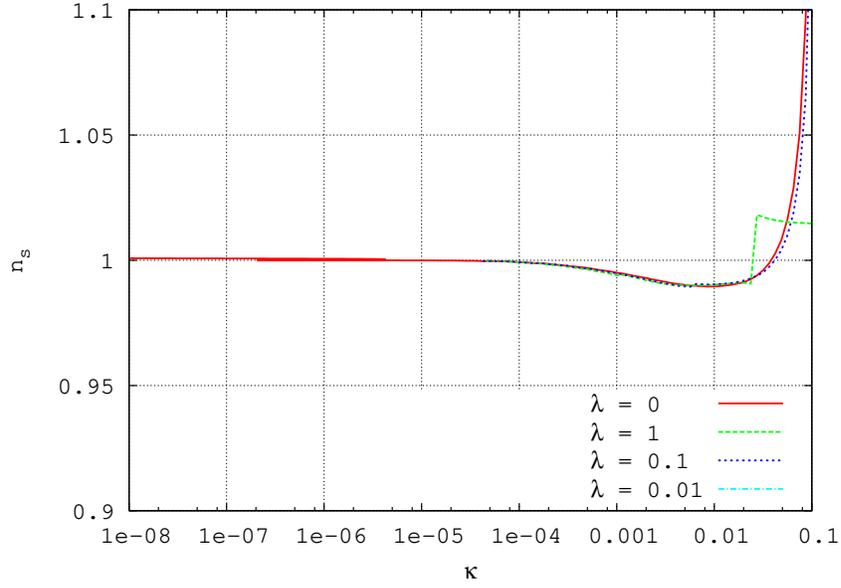}
\caption{$n$ vs. $\kappa$ for $\lambda =1,0.1,0.01,0$and $\mcN =1$ .}
\label{F:warmn}
\end{center}
\end{figure}

\begin{figure}
\begin{center}
\leavevmode\epsfysize=8cm \epsfbox{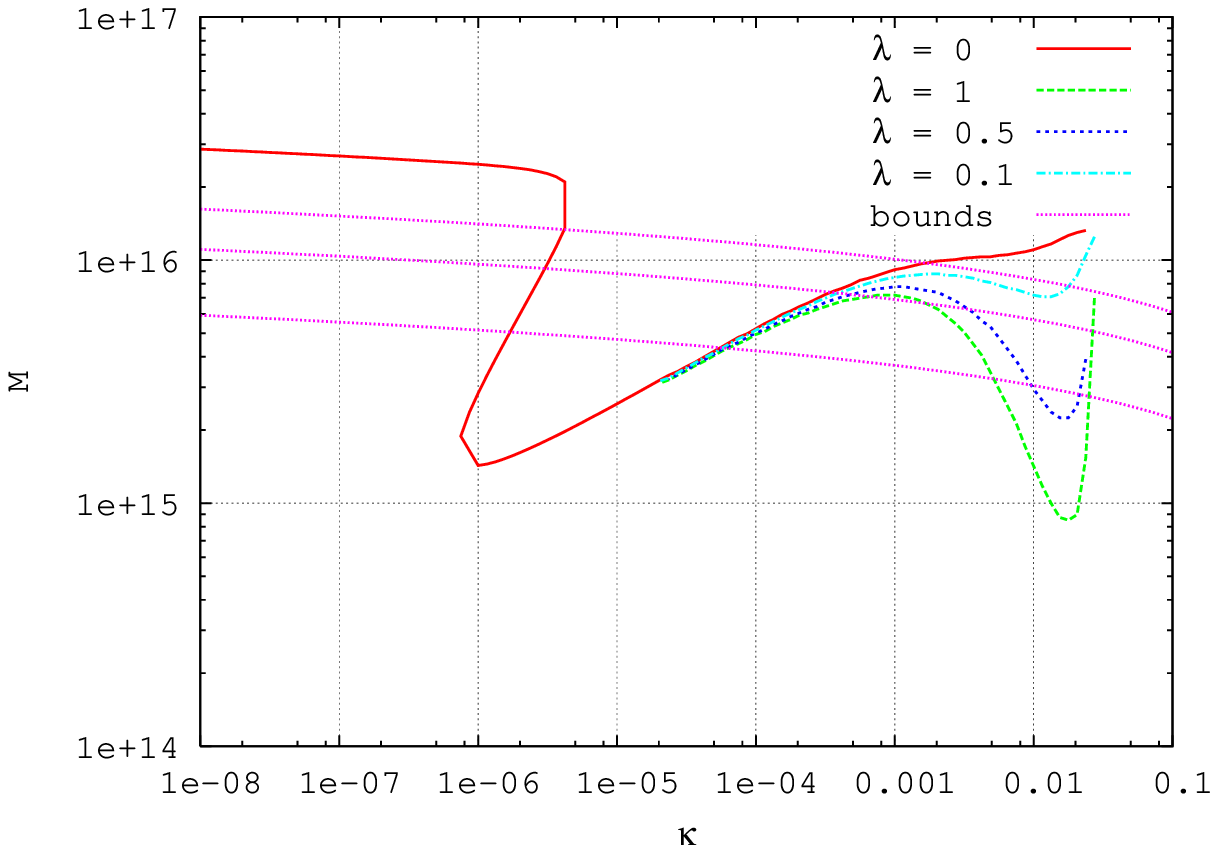}
\caption{$M$ vs. $\kappa$ for $\lambda =1,0.5,0.1,0$ and $\mcN
=16$. Also shown are, from top to bottom, the non-Gaussianity bound,
the 10\% bound, and the pulsar bound.}
\label{F:warm16}
\end{center}
\end{figure}

\begin{figure}
\begin{center}
\leavevmode\epsfysize=8cm \epsfbox{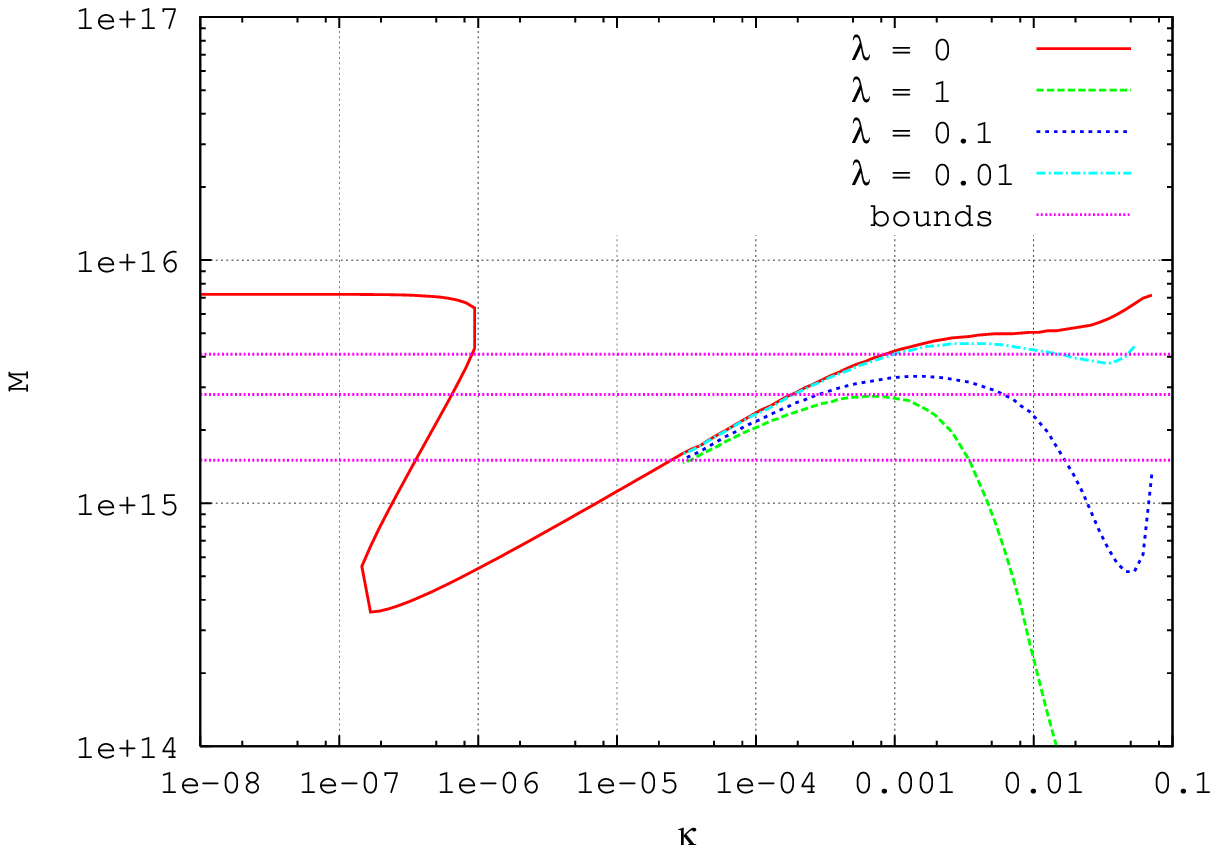}
\caption{$M$ vs. $\kappa$ for $\lambda =1,0.1,0.01,0$ and $\mcN
=1$. Also shown are, from top to bottom, the non-Gaussianity bound,
the 10\% bound, and the pulsar bound.}
\label{F:warmp}
\end{center}
\end{figure}

\section{Conclusions}

Cosmic strings form at the end of standard hybrid inflation. Both the
string and inflaton contribution to the CMB are proportional to the
same symmetry treaking scale $M$. This makes hybrid inflation testable via
CMB experiments. In this paper we have determined the parameter space
for which standard $F$-term hybrid inflation is compatible with the
CMB data. We considered both GUT $F$-term inflation and $P$-term
$F$-inflation.

We pointed out that cosmic strings forming at the end of GUT
$F$-inflation are not in the Bogomolny limit. The string tension
depends on the ratio of the masses of the Higgs and gauge fields, and
decreases in the limit of small quartic coupling constant.  This is why the
bounds on $M$ are relaxed compared to $P$-term models where the
strings do satisfy the Bogomolny bound.

We studied the inflationary scalar potential including all possible
soft and dissipative terms.  Supergravity corrections are calculated
assuming general expectation values in the hidden sector.  The
one-loop potential and the non-renormalisable terms are general, and
independent of the SUSY breaking mechanism in the true vacuum.  The
inflaton mass and linear $A$-term, on the other hand, depend
sensitively on the SUSY breaking mechanism.  These terms are large in
a canonical gravity mediated SUSY breaking scheme, in conflict with
the CMB data.  In order to evade the bounds, either these terms must
be tuned, or low energy SUSY breaking should take place after
inflation, or gauge mediation should be assumed.

Knowledge of the scalar potential of $F$-inflation allows for a
calculation of the perturbation spectrum.  We have derived analytical
formulas for the symmetry breaking scale $M$ as a function of the
superpotential coupling $\kappa$ in the limit that one term dominates
the potential.  These results agree well with our numerical
calculations.  The string tension, and thus the SSB scale $M$, is
bounded by the data. We used three different bounds, the 10\%
bound, the Kaiser-Stebbins bound and the pulsar bound (we believe the
Kaiser-Stebbins bound should be taken most seriously as the
theoretical uncertainties are smallest). For GUT strings, we find that
the relevant coupling $10^{-7}/\mcN \lesssim \kappa \lesssim
10^{-2}/\mcN$, with $\mcN$ the dimension of the Higgs-representation,
is still compatible with the data.\footnote{We note that although
Ref.~\cite{Rocher} considers GUT inflation, they do not take the
corrections away from the Bogomolny limit into account, and
consequently they find a much stronger bound}.  The bounds are stronger for the
strings formed in $P$-inflation: $10^{-7} < \kappa <10^{-4}$.

Finally we considered ways to ameliorate the CMB bounds.  In the
curvaton or inhomogeneous reheat scenario not the inflaton but some
other scalar field is responsible to the density perturbations.
Lowering the contribution of the inflaton to the CMB, even by only
10\%, immediately evades all bounds.  Warm inflation can occur if the
Higgs field is coupled to light fields with a large, order one,
coupling.  This opens up parameter space for large superpotential
couplings.

%The upcoming Planck satelite probes the level of non-Gaussianity to a
%much greater accuracy.  And either cosmic strings will be ruled out in
%standard $F$-term inflation, or their precense is detected via the KS
%effect.  A better resolution of the spectral index allows to
%discriminate between models of hybrid inflation with moderately large
%coupling.

\section*{Acknowledgements}
                                                                               
RJ would like to thank The Dutch Organisation for Scientific Research
[NWO] for financial support.

\end{document}